\documentclass{aa}  
\usepackage{graphicx}
\usepackage{txfonts}
\usepackage{float}

\usepackage{gensymb}
\usepackage{siunitx}
\usepackage{booktabs}

\usepackage{lipsum}
\usepackage{subcaption}       
\usepackage{lscape}            
\usepackage{placeins}
                                
\usepackage[colorlinks = true,
      linkcolor = blue,
      urlcolor = blue,
      citecolor = blue]{hyperref}

\begin{document}

   \title{Combining astrometry with pulsar timing: the first joint analysis of very low frequency gravitational waves}

   \subtitle{Astrometry and pulsar timing for low frequency GWs}

%
%
%

   \author{L. Filipello\inst{1, 2}\corrauth{lorenzo.filipello@unito.it}        
        \and W. Beordo\inst{1}
        \and V. Akhmetov\inst{1,3,4}
        \and M. Crosta\inst{1}
        \and M.G. Lattanzi\inst{1}
        }

 \institute{
INAF -- Osservatorio Astrofisico di Torino, Via Osservatorio 20, 10025 Pino Torinese (TO), Italy
\and
University of Torino, Department of Physics, Via Pietro Giuria 1, 10125 Torino, Italy
\and
Main Astronomical Observatory of the NAS of Ukraine, 27 Akademika Zabolotnoho St., 03143 Kyiv, Ukraine
\and 
Institute of Astronomy of V.N.Karazin Kharkiv National University, Svobody sq. 4, 61022 Kharkiv, Ukraine\\
}

 
\abstract
   {The pHz to sub-nHz gravitational wave (GW) regime remains largely unexplored but is crucial for mapping the early inspiral stage of supermassive black hole binaries (SMBHBs) and probing early-Universe physics. Astrometry and pulsar timing offer orthogonal and deeply complementary secular observables to investigate this frequency band.}
   {We aim to present the first joint data analysis combining real astrometric proper motions with binary pulsar timing, to search for and constrain continuous gravitational waves (CWs) sourced by SMBHBs in the ultra-low-frequency regime ($10^{-12}\,\text{Hz} \le f_{\text{GW}} \le 10^{-9}\,\text{Hz}$).} 
   {We employ a Bayesian model selection and upper limit estimation framework to combine apparent proper motion displacements of $\sim 1.5 \times 10^6$ quasars from the Gaia CRF3 catalog with the line-of-sight orbital period derivatives ($\dot{P}_b$) of 11 high-precision binary pulsars. To prevent spurious detections, we heavily model instrumental and astrophysical systematics: we propagate the Galactic potential uncertainty for pulsars via Monte Carlo simulations and perform a Vector Spherical Harmonics (VSH) decomposition up to the octupole order ($\ell=3$) for quasars.}
   {We find no statistically significant evidence for a CWs signal in the joint analysis ($\ln \mathcal{B}_{\text{joint}} = -0.42 \pm 0.03$). In the absence of detection, we set the tightest constraints to date on CW strain in the pHz band, yielding a 95\% upper limit of $h_0 \le 6.4 \times 10^{-11}$ at a reference frequency of $f_{\text{ref}} = 4 \times 10^{-10}\,\text{Hz}$. The combined dataset achieves full sky coverage and improves single-dataset upper limits by 20\%--30\%.}
   {Combining orthogonal observables successfully breaks spatial degeneracies intrinsic to isolated searches. Furthermore, forecasts from injection-recovery indicate that with the extended temporal baseline and reduced uncertainties of the upcoming Gaia DR4, this joint framework is poised to break the $h_0 < 10^{-11}$ upper limit barrier for sub-nHz CWs.}

   \maketitle

\nolinenumbers

\section{Introduction}
Thanks to observational detections of Gravitational Waves (GWs), new frontiers of physics have opened up in the last decades. 
However, while current cosmological models predict signals from mHz to sub-pHz frequencies, still no clear evidence for the presence of GW of any kind has been found in such a regime: claims of 3$\sigma$ detection for a Gravitational Waves Background (GWB) at $f_{ref} = 1yr^{-1}$ have been posed by PTA collaborations \citep{NANOGrav:2023gor}, approaching a confidence of 4$\sigma$ at the time of writing \citep{Taylor:2025lxy}. At the same time, the Planck experiment was only able to put an upper limit on inflation-driven GWB at $\sim 10^{-18}$ Hz \citep{Pagano:2015hma}.

Nevertheless, even with the combination of all these experiments, a broad range, namely the pHz region, is still a quite unexplored, yet crucial, observational regime. Gravitational waves in this band are produced by two main classes of astrophysical and cosmological phenomena. The first is represented by extremely massive outliers Supermassive Black Holes Binaries (SMBHBs) in the early stage of inspiral, with periods spanning from years to millenia. These sources will produce quasi-monochromatic continuous signals. The second kind of GW is a Stochastic Gravitational Wave Background (SGWB) of cosmological origin: in the sub-$\text{nHz}$ to $\text{pHz}$ regime, a cosmological background can arise from different early-Universe processes. These include cosmic strings \citep{Wachter:2024zly}, where radiating loops produce a broad-band stochastic signal extending to very low frequencies, phase transitions at intermediate energy scales \citep{Bagherian:2025puf}, and primordial metric perturbations associated with inflation or primordial black hole formation scenarios \citep{Bi:2026zlt}. Probing this frequency band allows us to map the high-mass tail of the SMBHB population and place independent constraints on early-Universe physics, with astrometry as one of the main candidates for filling this gap.

The current state-of-the-art experiment for astrometry is Gaia \citep{Gaia:2016zol}, whose goal is the micro-arcsecond precision for catalog astrometric parameters.
It has been shown how a passing GW can perturb the incoming photons from celestial objects, resulting in an apparent and periodic motion in the sky \citep{Pyne:1995iy}. Even though some works have tried to define possible data analysis pipelines for GW periodic signals using Gaia time series \citep{Moore:2017ity, Geyer:2024lvk}, this instrument was not designed to provide astrometric time series (epoch astrometry will only be published with the fourth Data Release); therefore, all current constraints from real astrometric data are derived from proper motions \citep{Akhmetov:2026bfn, Darling:2024myz, Jaraba:2023djs}.

Using astrometric proper motions as a GW induced secular effect, we can in principle observe any GW signal with a period larger than the total observation time. To implement our first joint analysis, we use the Gaia Celestial Reference Frame (CRF) catalog of quasars from DR3, or Quasi Stellar Object (QSO) \citep{Gaia:2022huk}, due to the high precision and high number of sources in the dataset.
While the standard Pulsar Timing technique focuses on periodic signals confined between the Nyquist limit and the observation time of the experiment \citep{NANOGrav:2023hde}, different approaches have been proposed to explore pHz signals using secular effects of pulsar binaries \citep{Kumamoto:2019uoy, DeRocco:2023qae, Zheng:2025tcm}. 

Due to their strong complementarity, a cross-correlation between these two different and independent measurements could enhance the sensitivity we can achieve on low-frequencies GW  \citep{Cruz:2024diu}, with optimistic forecasts of joint-analysis reaching an improvement factor of 50\% compared to Pulsar Timing observations alone \citep{Perna:2026zww}. However, no combined analysis on real datasets has ever been made until now. With this work, we develop a data analysis pipeline for searching and characterizing a continuous gravitational waves (CWs) sourced by a single Supermassive Black Hole Binary, as very loud and slowly rotating SMBHBs are one of the most prominent candidates for GW sources in this regime. Nevertheless, the analysis pipeline can easily be implemented for a broader range of sources.

In the following section we define and briefly explain the formalism of the GW effects on the two different observables, altogether with all other astrophysical and systematic expected contributions. Then, in Sec. \ref{sec:dataset}, we present the used datasets and describe how they were constructed. Moving to Sec. \ref{sec:data_analysis}, the whole data analysis framework and pipeline is defined. Finally, in Sec. \ref{sec:results} and \ref{sec:future}, we show the obtained results and the improvements we expect in the (nearby) future.

\section{Theoretical Framework}\label{sec:theo_frame}
In this section, we highlight and summarise the theoretical derivation of the formulas used in the following analysis. Here, we combine two different, yet complementary, effects caused by a passing GW. Since we are confining ourselves to secular effects, rather than periodic signals, the GW periods range we can measure spans from the look-back time (lowest frequency limit) to the total observation period (highest frequency limit). Moreover, since we also restrict the analysis to CW, a tighter limit on observable frequencies is set by loss of coherence due to the distance horizon limit, putting the lowest limit to roughly $10^{-12}$ Hz \citep{Margalit:2020sxp}. \\
In order to explain the actual observables, we define the GW notation; for a Single Source (SS) GW, the perturbation is written as:
\begin{equation}
    h_{ij}(t,\hat{x}) = e_{ij}^+ (\hat k)  h^+(t,\hat x) + e_{ij}^{\times} (\hat k) h^{\times}(t,\hat x)  \; ,
\end{equation}
with
\begin{align}
    h^+ ={} & h_0 \cos(2 \Psi) (1 + \cos^2(\iota)) \sin(2 \phi) \\ \nonumber
    &+ 2 h_0 \sin(2\Psi) \cos(\iota) \cos(2\phi) \;, \\
    h^{\times} ={} & h_0 \sin(2 \Psi) (1 + \cos^2(\iota)) \cos(2 \phi) \\ \nonumber &+ 2 h_0 \cos(2\Psi) \cos(\iota) \sin(2\phi) \;.
\end{align}
Here, $h_0$ is the strain of the GW, $\iota$ and $\Psi$ are the inclination and polarization angles of the GW source, $\hat k (\alpha_{GW}, \delta_{GW})$ is the normalised propagation direction of the wave and
\begin{equation}
    \phi = \omega (t - \hat k \cdot \hat x) + \Phi_0 \;,
\end{equation}
with $\omega$ as the angular velocity and $\Phi_0$ as the phase. Assuming that such GW is formed by a SMBHB, the amplitude of the strain is related to the chirp mass $\mathcal{M}$ and luminosity distance $d_L$ through
\begin{equation}
    h_0 = \omega^{2/3} \dfrac{2(G \mathcal{M})^{5/3}}{d_{L} c^4} \;.
\end{equation}
Also the tensor bases $e_{ij}^+$ and $e_{ij}^{\times}$ can be expressed in terms of the GW source sky direction $\hat \alpha_{GW}$ and $\hat \delta_{GW}$:
\begin{equation}
    \begin{cases}
    \hat \alpha_{GW} = (- \sin(\alpha_{GW}), \cos(\alpha_{GW}), 0) \;, \\
    \hat \delta_{GW} = (-\sin(\delta_{GW}) \cos (\alpha_{GW}),- \sin(\delta_{GW}) \sin (\alpha_{GW}),  \cos (\delta_{GW})) \;. 
    \end{cases}
\end{equation}
as
\begin{equation}
    \begin{cases}
        e_+ = \hat \alpha_{GW} \hat \alpha_{GW} - \hat \delta_{GW} \hat \delta_{GW} \;, \\
        e_\times = - (\hat \alpha_{GW} \hat \delta_{GW} + \hat \delta_{GW} \hat \alpha_{GW}) \;.
    \end{cases}
\end{equation}
Finally, since we restrict our analysis to GW periods much larger than observation time, i.e.  $\omega t << 1$, the following approximation will hold:
\begin{equation}
     \phi \simeq -\hat k \cdot \hat x + \Phi_0 \;.
 \end{equation}

 \subsection{Radial Effect}
In the context of Pulsar Timing, GW can modify the time of arrival of millisecond pulsars photons, from which we can infer the GW parameters. However, when dealing with periods longer than observation time, a GW can induce an apparent velocity between the pulsar and the observer \citep{Romano:2016dpx}, that can be written as
\begin{equation}\label{eq:gw_velocity}
    v_{GW} =\sum_{a=+, \times} \dfrac{\hat n_p^i \hat n_p^j e_{ij}^a(\hat k) }{2(1 + \hat k \cdot \hat n_p )} (h_a(t,0) - h_a(t-d_p, \hat d_p)) \; ,
\end{equation}
where $\hat d_p$ is the pulsar distance vector and $d_p$ its magnitude. The time derivative of this equation represents the GW induced acceleration $a_{GW}$. Therefore, under the effect of a GW, the time derivative of a periodic signal will be directly proportional to the GW induced acceleration and the unperturbed period:
\begin{equation}
    \dot P = a_{GW} P \;,
\end{equation}
where the complete formula is provided in eq. \eqref{eq:gw_acc} in Appendix B. \\
For this analysis we will use pulsar binaries and their orbital period $P_b$ parameters, due to the high precision and stability of the measurement through all PTA collaborations. \\
Low-frequencies GW are not the only contributions to orbital period changes, since other astrophysical effects are present. The most relevant are the following (more negligible going right):
\begin{equation}
    \dfrac{\dot{P}_{b, \text{obs}}}{P_{b}} = a_{shk} + \dfrac{\dot{P}_{b, \text{gal}}}{P_{b}} +\dfrac{\dot{P}_{b, \text{int}}}{P_{b}}  + a_{GW}
\end{equation}
The first term is the Shklovskii kinematic effect \citep{Shklovskii1970}, i.e. the Doppler effect due to the angular proper motion $\mu$ of a celestial object. Its magnitude is $a_{shk} = \mu^2 d_p / c$. \\
The second term is the orbital change due to the Milky Way gravitational potential induced acceleration. The contribution to the observed $\dot P_b$ is the line of sight net acceleration: since we also experience the Galactic potential, we can only measure the net acceleration
\begin{equation}
    \dfrac{\dot{P}_{b, \text{gal}}}{P_{b}} = \dfrac{(a_p - a_{SS}) \cdot \hat{n}_p}{c} \;,
\end{equation}
where $a_{SS}$ is the Galactic acceleration at Solar System position and $a_p$ is the acceleration at pulsar position. The first one can be directly measured \citep{GaiaEDR3_Acceleration_2021}, while the second can be estimated from Galactic models. \\
The last term to define is the intrinsic orbital deceleration: a compact binary will lose energy due to the continuous emission of GW, resulting in a change of the fractional time derivative of the orbital period \citep{Peters:1963ux}:
\begin{equation}
    \dfrac{\dot{P}_{b, \text{int}}}{P_{b}} = -\dfrac{192 \pi G^{5/3} f(e)}{5c^5} \left( \dfrac{P_b}{2 \pi} \right)^{-8/3} \dfrac{m_p m_c}{(m_p + m_c)^{1/3}} \;,
\end{equation}
where $m_p$ is the pulsar mass, while $m_c$ is the mass of the companion, and $f(e)$ is a function of the orbital eccentricity $e$
\begin{equation}
    f(e) = \left(1 + \dfrac{73}{24}e^2 + \dfrac{37}{96} e^4\right)  (1 - e^2)^{-7/2}.
\end{equation}
In absence of more accurate measurement, we assume the mass of the pulsar to be the standard $1.4 \ M_{\odot}$ and use the inferred median for the companion mass.

\subsection{Tangential Effect}

We will now describe the complementary effect measured by astrometry. It has been shown that a passing GW can perturb the path of a photon emitted from a celestial object, resulting in an apparent position displacement. In the far-away zone approximation, with respect to a static observer tetrad, this effect has an analytical formula \citep{Pyne:1995iy}:
\begin{equation}\label{gw_displacement}
      \delta n^i_{GW}  = \sum_{a=+,\times} \left[ \frac{(\hat n^i + \hat k^i) h_a(t,0)}{2 (1 + \hat{n} \cdot \hat{k})}e_{jk}^a(\hat k) \hat n^j \hat n^k - \frac{1}{2} e_{ij}^a(\hat k) \hat n^j h_a(t,0)\right] \;,
\end{equation}
where $\hat k$ is the GW direction of propagation and $\hat n$ is the unperturbed position of the celestial object on the sky sphere. Due to the extreme distance of quasars, the "star" term is neglected and just treated as random white noise.\\
Even in this scenario, if we restrict to frequencies smaller than the inverse of observation period, only secular effects shall survive. For the astrometric displacement, this secular effect is an apparent proper motion $\mu_{GW}$, i.e. the time derivative of eq. \eqref{gw_displacement}. For a generic celestial object, the measured proper motion is a combination of several factors:
\begin{equation}
    \mu_{obs} = \mu_{\text{int}} + \mu_{\text{glide}} + \mu_{\text{sys}}^* + \mu_{\text{GW}} \; ,
\end{equation}
where the first term is the intrinsic proper motion of the object, potentially caused by Galactic gravitational potential, local interactions, stellar streams or supernova ejections, or transient events.\\
The second term is the so-called "glide", or secular aberration, namely the apparent motion of stars we see due to the free-fall of the Solar System toward the Galactic centre:
\begin{equation}
    \mu_{glide} = \frac{1}{c} \left[  a_{SS} - ( a_{SS} \cdot \hat n) \hat n \right] \;.
\end{equation}
We define the third term as $\mu_{\text{sys}}^* \equiv \mu_{\text{rot}} + \mu_{\text{sys}}$, where $\mu_{\text{rot}} $ is the apparent proper motion due the reference frame's rotation (or rotation of the universe), while the last term will collect all unknown systematic effects. We highlighted 
$\mu_{glide}$ and $\mu_{rot}$ since their effect is purely a dipolar contribution, hence they can be precisely estimated from a Vector Spherical Harmonics (VSH) decomposition. \\
For Gaia stars, $\mu_{\text{int}}$ is typically orders of magnitude greater than other contributions. However, due to the cosmological distance of quasars, their intrinsic proper motion is absolutely below the instrument precision: this means that any non-zero measured proper motion for QSOs is purely caused by the astrophysical and systematic effects explained above.

\begin{figure*}
    \centering
    \includegraphics[width=0.9\linewidth]{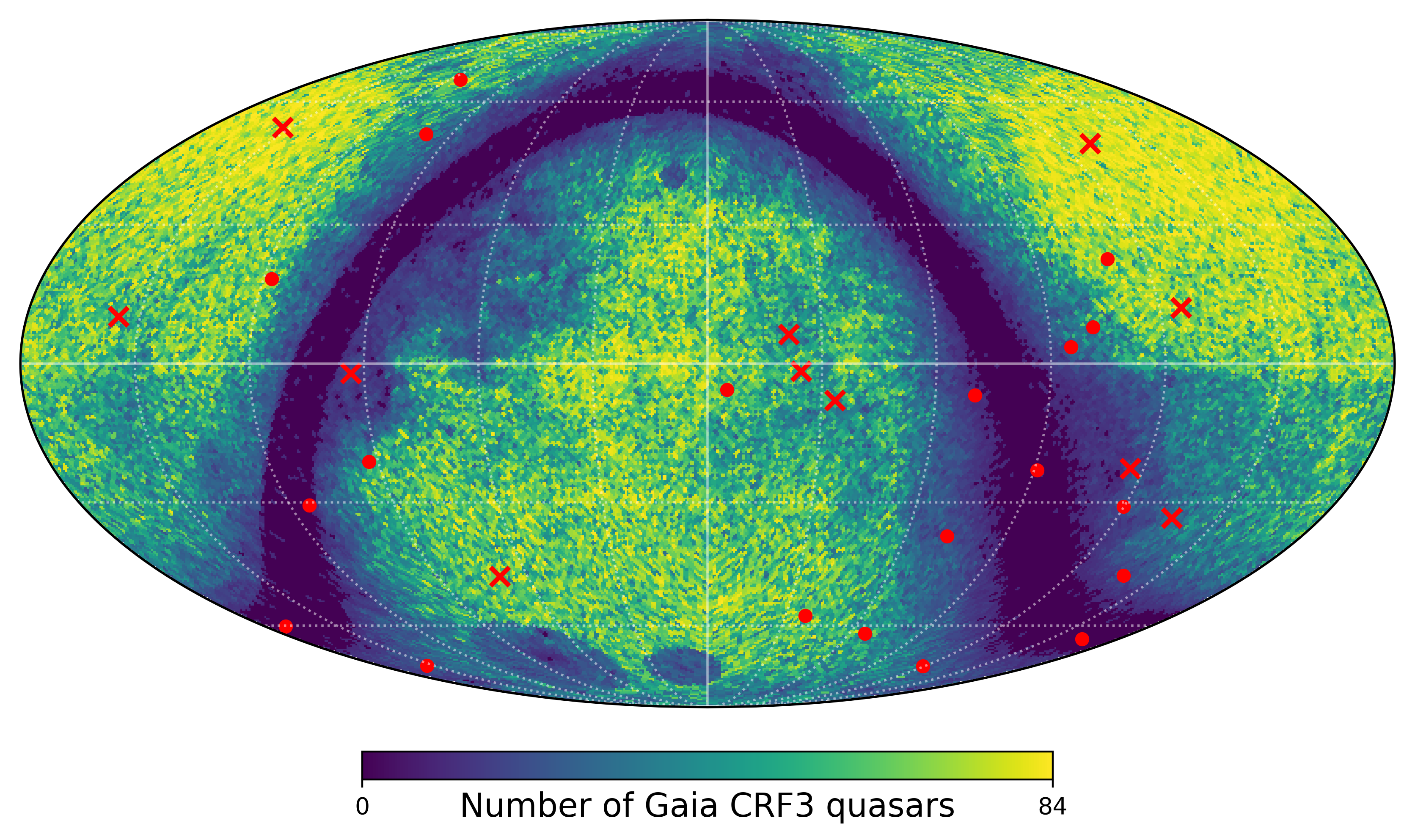}
    \caption{Spatial density of QSOs from the Gaia DR3 catalog and positions of all pulsar binaries in the ATNF catalog (red markers). The red crosses are the pulsars satisfying the imposed mask and therefore used for the analysis.}
    \label{fig:qsoxpta_positions}
\end{figure*}
\section{Dataset}\label{sec:dataset}

\subsection{Gaia CRF3}

The Gaia CRF3 catalog provides astrometric parameters of $\sim 1.6$ millions of QSOs. It is the result of Gaia Data Release 3, comprising 34 months of observations (from July 2014 to May 2017). The majority of these quasars are at a redshift of $z \approx 1.5 - 2$, putting the accessible frequency range between $10^{-17} \ \text{Hz} < f_{GW} < 5.6\times10^{-9} \ \text{Hz}$ (note that for the CW analysis we set the effective lower limit to 1 pHz for loss of coherence at lower frequencies \citep{Margalit:2020sxp}). \\
While there are other QSO catalogs with higher purity such as Quaia \citep{Storey-Fisher:2023gca}, we decide to use Gaia CRF3 because it also provides proper motion components covariance. Neglecting the correlation between $\mu_{\alpha}$ and $\mu_\delta$ would result in two downsides: the most straightforward is underestimating the total errors (and therefore the upper limits we can set), and secondly, all ellipsis' uncertainties will have the same orientation, potentially biasing the results of a CW analysis towards a preferred GW direction of propagation \citep{Akhmetov:2026bfn}.

We only exclude sources with goodness-of-fit parameter $\texttt{RUWE} \leq 1.4$ and total proper motion amplitude above $1 $ mas/yr, to exclude possible contaminations from stellar sources or astrometric artifacts. This results in keeping in the analysis also objects with very high proper motions (almost $\sim$400k sources with $\mu > 0.5 \ \text{mas/yr}$), but we preferred to improve the modelling of signals rather than dealing with a selection bias on the resulting upper limits. \\
\begin{table}[h!]
\caption{Statistical properties of the 1\,505\,427 quasars from the Gaia CRF3 catalog.}
\label{tab:quasar_stats}
\begin{tabular}{l|c|c|c|c}
\hline\hline
\textbf{Parameter} & \textbf{Mean} & \textbf{St. Dev.} & \textbf{Min.} & \textbf{Max.}  \\
\hline
$\mu^*_\alpha$ ($\mu$as~yr$^{-1}$) & $-0.68$ & 793     & $-8\,521$ & 8\,728\\
$\mu_\delta$ ($\mu$as~yr$^{-1}$) & $-1.37$ & 736 & $-8\,566$ & 9\,297 \\
$\sigma_{\mu^*_\alpha}$ ($\mu$as~yr$^{-1}$) & 653 & 476 & 9.72   & 3\,365 \\
$\sigma_{\mu_\delta}$ ($\mu$as~yr$^{-1}$) & 606 & 441 & 12.98  & 3\,387 \\
\hline
\end{tabular}
\end{table}
Due to high computational costs, an isolatitude pixellization is performed on the catalog, as it has been shown that this kind of data compression can bring substantial boosts in time performance at a negligible cost in information loss. For each pixel we use the weighted (through inverse of variance) value of proper motions components, with Gaussian errors of the mean as pixel's proper motion components uncertainties. Also mean correlation of $\mu_\alpha$ and $\mu_\delta$ is taken into account for the caclulations of the full covariance matrix of each pixel.

Pixels with less than 10 objects are not included in the analysis, in order to avoid portions of the sky with fewer sources to have a major impact on the final results.

\subsection{Pulsar Binaries}

We construct our pulsar binaries dataset from the ATNF Pulsar catalog  \citep{Hobbs:2003gk} after applying careful restrictions. The major issue with this kind of analysis is related to the binary distance estimation. Almost every contribution to $\dot P_b$ heavily depends on the distance, namely, the leading order correction $a_{shk}$, the Galactic acceleration, and the Pulsar term of $a_{GW}$. For this reason, we only use binaries with well-defined parallax $\varpi$ measurements and with $\sigma_\varpi / \varpi < 10\%$. Moreover, we also apply a cut over the distance to alleviate the degeneracies related to phase multiplicity of the GWs and diminish the uncertainties on the Galactic acceleration; therefore, we only include binaries closer than 1 kpc. \\
\indent This results in a final dataset of 11 binaries with good estimates of distance, proper motion, position, and companion mass. Unfortunately, due to Interstellar Medium extinction, most of the binaries are located outside the stellar disk, therefore the isotropy of the joint dataset does not improve significantly. This issue should drastically improve with SKA telescope, which is supposed to see and investigate the millisecond pulsar population close to the Galactic bulge \citep{Braun:2015zta}. \\
\indent A table with the pulsars used in the analysis and their parameters is provided in Appendix. Moreover, Fig. \ref{fig:pta_contributions} shows the observed values of $\dot P_b / P_b$ compared to all astrophysical contributions. As expected, the kinetic term is the predominant one for the majority of the binary systems except for B1534+12, where the dominant term is the intrinsic emission due to several factors (massive neutron star as companion, high eccentricity, short orbital period). \\
Finally, all line of sight Galactic acceleration have a negative sign (except for J0613-0200 and J1614-2230), indicating a deceleration towards the Sun.

In fig. \ref{fig:qsoxpta_positions} we present the spatial distribution of the combined datasets used in this analysis. It is possible to visualise the extinction from the Galaxy disk and bulge affecting the number of quasars observed by Gaia.

\section{Data Analysis}\label{sec:data_analysis}

In this work, we adopt a Bayesian framework for statistically robust signal detection and upper limit estimation. After defining the likelihoods for the problem, we adopt a Nested Sampling algorithm for signal detection and a Monte Carlo Markov Chain (MCMC) for the calculation of the posterior distributions.\\
For a CW search, we need 7 parameters to model the GW (assuming such a wave is emitted by a SMBHB), i.e. the strain $h_0$, the GW frequency $f_{GW}$, the direction of propagation ($\alpha_{GW}, \delta_{GW}$, in ICRS coordinates), the orientation angles of the SMBHB $\iota$ and $\Psi$ and the random phase $\Phi_0$. We define a standard Gaussian likelihood for pulsar search: 

\begin{equation}
    \mathcal{L}_{PTA} (\Theta_{GW},d_n) = \prod_{n=1}^{N_{pul}} \dfrac{1}{\sqrt{2 \pi} \sigma_n} \exp \left[- \dfrac{\text{res}^2_n(\Theta_{GW},d_n)}{2 \sigma_n^2} \right] \; ,
\end{equation}
where $\text{res}_n = \dot P_{b, exp} / P_b - \dot P_{b, obs}/P_b$ is the residual of the $n$-pulsar, $d_n$ is the pulsar distance, $\sigma_n$ is the total uncertainty on the residual, and $\dot P_{b, exp}$ is the derivative of the modelled period. Even though we only selected binaries with good parallax uncertainties, all contributions (except for the intrinsic emission) can assume significantly different values within the error range. Therefore, in order to reduce the dimension of the parameter space of the MCMC, we fix position and proper motions (and propagate their errors in the total uncertainties), as they are not the main source of uncertainty, while keeping the distance as a free parameter.

The Galactic acceleration is the most complicated term since it is heavily model-dependent and there is no consensus neither on the total mass of the Galaxy nor on its components \citep{ouDarkMatterProfile2024, beordoExploringMilkyWay2024, jiaoDetectionKeplerianDecline2023, zhouCircularVelocityCurve2023, eilersCircularVelocityCurve2019, McMillan2017}. Therefore, estimating the uncertainty of the Galactic acceleration is key in order to avoid spurious detections. Despite the recent claims significantly reweighting the Galaxy virial mass down to $2 \times 10^{11} \, \mathrm{M}_\odot$ \citep{ouDarkMatterProfile2024, jiaoDetectionKeplerianDecline2023}, in this work we refer to the standard McMillan17 model \citep{McMillan2017}. 
Given this model, we assess how the Galactic acceleration changes with varying the model parameters over their posterior distributions. 
In Appendix~\ref{sec:gal_model}, we detail the Monte Carlo approach used to compute the uncertainty on the Galactic acceleration, given the uncertainties on the model parameters. Then, for each pulsar binary, we add this model-induced variance to the total residual variances $\sigma_n^2$. \\
We stress the fact that without this additional uncertainty the residuals differ significantly from zero, as shown in \citep{Moran:2023myv}, were a jitter parameter is defined to absorb the excessive power, meaning that the expected values do not sufficiently describe the residuals or that the observational error is underestimating the actual uncertainty. Due to the high precision and accuracy of astrometric parameters and without raw data of timing parameters, we infer that this extra noise should come from the highly uncertain Galactic modelling.

A similar approach is applied for the analysis of quasars. Since proper motions are two-dimensional vectors, we construct a Gaussian likelihood from the standard Mahalanobis distance:

\begin{equation}
    \mathcal{L}_{Gaia} = \prod_{p=1}^{N_{QSO}} \dfrac{1}{ 2 \pi\sqrt{ \det (C_p)}}\exp \left[ -\dfrac{1}{2}\text{res}_p^T C_p^{-1} \text{res}_p\right] \;,
\end{equation}
where $res_p = (\mu_{\alpha,obs} - \mu_{\alpha,exp} , \mu_{\delta,obs} -\mu_{\delta,exp})$ and $\mu_{exp}$ takes into account systematic contributions and GW effect. $C_p$ is the $2 \times2 $ covariance matrix for each quasar, containing errors of proper motion components and their correlation. We find that dipole contribution alone is not sufficient for describing quasars proper motion power, as Gaia CRF3 has non-zero systematic correlations at medium and small scales \citep{Gaia:2022huk}. As a consequence, the algorithm forces the GW parameters to absorb this residual power, resulting in spurious posteriors of the GW strain. As shown in Appendix \ref{sec:gaia_systematics}, after implementing a harmonic decomposition, a plane GW should exhibit its signal mainly at quadrupolar level, with smaller fractions at each higher order. Therefore, we propose a hybrid technique implementing standard Bayesian inference and Vector Spherical Harmonics \citep{Mignard:2012xm} analysis: besides dipole contributions, VSH coefficients are also implemented as free parameters in the MCMC. \\
We cut the VSH analysis at the $\ell \leq 3$ level, i.e. only quadrupole and octupole contributions are implemented, since the higher computational cost was not resulting in an enhanced GW constraining performance. Therefore, the final value of $\mu_{exp}$ (and hence the astrometric likelihood) will be a function of GW parameters $\Theta_{GW}$ and VSH coefficients $\Theta_{VSH}$. \\
A proper definition of the VSH decomposition employed, its justification and the astrometric-only results are fully discussed in Appendix \ref{sec:gaia_systematics}.\\
Finally, we have all the building blocks for the definition of our joint analysis likelihood:
\begin{equation}
    \mathcal{L}_{joint} = \mathcal{L}_{PTA} \times\mathcal{L}_{Gaia} \;,
\end{equation}
which will be a function of GW parameters, VSH parameters and pulsar distances.

\subsection{Priors}

According to Bayes theorem, the formulation of the posterior distribution, requires specific likelihood function and prior distribution: in the context of setting upper limits, Bayesian evidence serves exclusively as a normalization constant, leaving the functional form of the posterior unaltered. In the absence of any previous detection, we adopt isotropic uniform priors for the direction of propagation $(\alpha_{GW}, \delta_{GW})$ and source binary orientation $(\iota, \Psi)$. We choose uniform prior distributions for strain $h_0$ and random phase $\Phi_0$. Finally, GW frequency is sampled from a log-uniform distribution, since our analysis spans over several orders of magnitude. Moreover, frequency boundaries are carefully selected in order to avoid physically inaccessible periods.

Regarding quasars glide and rotation dipolar contributions, we cannot use Gaia CRF3 results for the prior distributions, since our analysis is constructed on the same dataset. These data double-counting could, in best scenario, underestimate the real uncertainties on the free parameters or enforce possible bias and systematics present in the catalog. Therefeore, as nominal value, we choose an independent VLBI measurement \citep{Titov_2018} of the secular drift of magnitude $5.2 \mu \text{as / yr}$ directed towards the Galactic centre $\alpha_{glide} = 266 \degree $ and $\delta_{glide} = 29 \degree$. Since we assume the cosmological principle of a non-rotating universe, we set a zero rotation vector as nominal value. For both dipole contributions we choose a Gaussian prior distribution with a loose standard deviation of $5 \mu \textit{as /yr}$, while we set uninformative uniform priors on the remaining VSH coefficients. 

The final posterior distribution will be the product of the joint likelihood and the joint prior distribution $ \pi = \prod_{i = 0}^{N} \pi(\Theta_i)$.

\section{Results}\label{sec:results}

\begin{table}[]
    \centering
    \begin{tabular}{c|c|c}
    \hline \hline
        \textbf{Parameter} & \textbf{Distribution} & \textbf{Domain} \\
        \hline \hline
        $h_0$ & uniform & $[10^{-20} , 10^{-4}]$ \\
        $\log f_{GW}$ & uniform & $[-12 , -9.4]$ \\
        $\alpha_{GW}$ & uniform & $[0, 2\pi)$ \\
        $\delta_{GW}$ & $\frac{1}{2} \cos \delta_{GW}$ & $(-\pi , \pi)$ \\
        $\Psi$ & uniform & $[0 ,2 \pi)$ \\
        $\iota$ & $\frac{1}{2} \cos \iota$ & $(-\pi , \pi)$ \\
        $\Phi_0$ & uniform & $[0, 2\pi)$ \\
        \hline \hline

    \end{tabular}
    \caption{GW priors distributions and allowed ranges.}
    \label{tab:gw_priora}
\end{table}

\subsection{Signal Detection}

We choose a Dynamic Nested Sampling algorithm from \texttt{dynesty} library for the calculation of the Bayes factor (BF) defined as 
\begin{equation}
    \mathcal{B} = \dfrac{Z_1 = \int \mathcal{L}(\Theta_1) \pi(\Theta_1) d\Theta_1}{Z_0 = \int \mathcal{L}(\Theta_0) \pi(\Theta_0) d\Theta_0},
\end{equation}
where $Z_0$ is the evidence for the null hypothesis, hence no GW effects (only astrophysical or other instrumental systematics), while $Z_1$ is the evidence for a GW signal. In scenarios like ours, i.e. SNR$=0$, high auto-correlations of parameters and heavy systematic effects, nested sampling algorithms perform significantly better and are more robust than standard MCMC methods, which easily get trapped in local extrema or fail to explore degenerated parameter spaces \citep{Buchner:2021kpm}. 

Starting from the pulsar binaries dataset, we find similar results to \citet{Zheng:2025tcm}, namely:
\begin{equation}
    \ln \mathcal{B}_{PTA} = -2.82 \pm 0.41 \;,
\end{equation}
indicating a strong preference for the model with no GW signal. Repeating the analysis with mock signal injections showed that the False Positive Rate (FPR) of the analysis is well bounded, with no log-Bayes factor $>-1$.

For the astrometric search, due to the presence of non negligible systematic effects, we define three different scenarios: 
\begin{itemize}
    \item $M_0$ only dipole contribution,
    \item $M_1$ dipole, quadrupole, and octupole contributions,
    \item $M_2$ dipole, quadrupole, octupole contributions, and CWs signal,
\end{itemize}
and estimate the evidence for each proposed model. \\
We find the following results:
\begin{itemize}
    \item $\ln \mathcal{B}_{10} = \dfrac{Z_1}{Z_0} = 14.42 \pm 0.40$ \;, \\
    \item  $\ln \mathcal{B}_{21} = \dfrac{Z_2}{Z_1} = 1.41 \pm 0.52$ \;,
\end{itemize}
where $Z_i$ is the evidence for model $M_i$. The retrieved values show a very strong evidence, according to the Kass-Raftery scale \citep{Kass:1995loi}, for the model with $\ell = 2,3$ contributions compared to the one with just dipole effects. This indicates the fact that the secular aberration and rotation are not sufficient to explain the quasars' proper motions, which is in total agreement with results in literature \citep{Gaia:2022huk}. On the other hand, $\ln \mathcal{B}_{21} > 1$ shows a weak preference for the model with a GW; however, there are few aspects to consider: this value is $3\sigma$ compatible with zero and the very high-dimensionality of the model can bring numerical noise (in the $h_0 \to 0$ regime, the remaining parameter posterior distribution is completely flat and uninformative). As we show in Appendix \ref{sec:gaia_systematics}, in the full Gaia CRF3 catalog, there is a residual power also at $\ell>3$: as a result, the Nested Sampling algorithm is weakly overfitting this excessive power at high multipoles. \\
Instead of adding even more parameters to the model, we use the fact that $M_1$ is a nested model of $M_2$ in the limit $h_0 \to 0$ and evaluate the Bayes factor using the Savage-Dickey Density Ratio (SDDR) as a countercheck. The new BF is defined as the ratio between the posterior density at $h_0 = 0$ and the prior density in the same scenario:
\begin{equation}
    \mathcal{B}_{SDDR} = \dfrac{p\left(\Theta=\Theta_0|y, M_2\right)}{p\left(\Theta=\Theta_0|M_2\right)} \;.
\end{equation}
We find that, using SDDR method for scenario $M_2/M_1$ (i.e. GW signal vs. no GW signal), $\ln \mathcal{B}_{SDDR} = -0.71 \pm 0.03$, which indicates no evidence for a CWs signal. We conclude that also in the astrometric catalog there is no robust evidence of a GW detection and therefore move on to the joint analysis.

For the combined Bayes factor, due to the above discussed problems, we stick to the SDDR method. The model $M_1$ will consider previously discussed astrophysical effects for pulsar binaries and generic VSH contributions up to octupole order for quasars plus a coherent GW signal. The nested model is defined as $M_1$ in the limit $h_0 \to 0$. We find the following value:
\begin{equation}
     \ln \mathcal{B}_{joint} =   -0.42 \pm 0.03 \;,
\end{equation}
showing no evidence for a GW signal even in the combined datasets. We assess the FPR by sky scrambling: while keeping the real values of $\mu$ and $\dot P_b/P_b$ with their corresponding errors, we randomly shuffle the position of the objects in the two datasets. After creating 50 samples, none of them exceeded the value of $\ln \mathcal{B} = -0.2$, with all realisations being more than 5$\sigma$ from threshold value of zero. We conclude that our analysis is well-bounded with minimal risks of False Positive detection. \\
\begin{figure*}[h]
    \centering
    \includegraphics[width=0.95\linewidth]{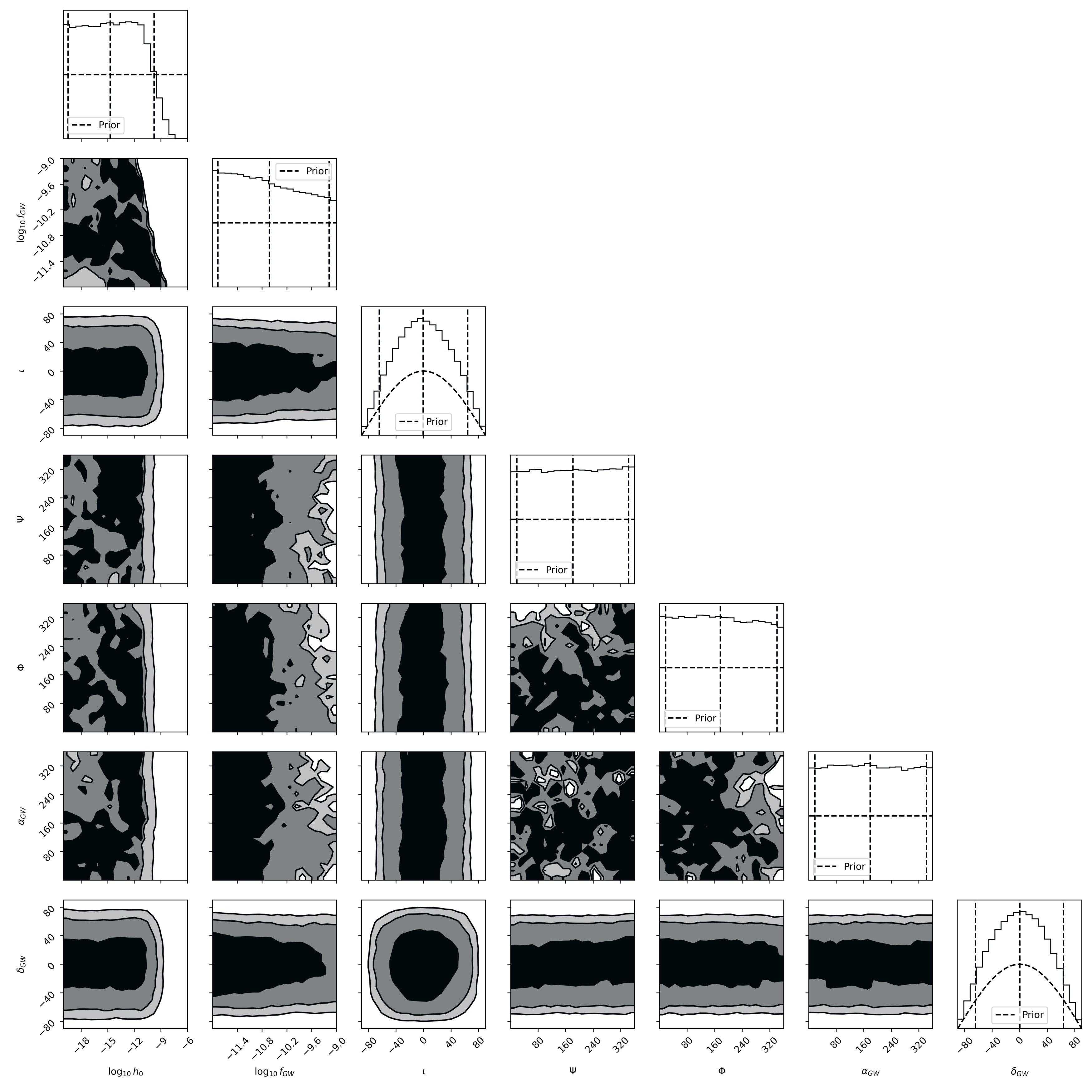}
    \caption{Posterior distribution of the joint analysis. The shaded areas represent, from darker to brighter, the 50-th, the 88-th and 95-th quantiles. The black dotted lines show qualitatively the adopted prior. The GW strain was sampled with log-uniform prior distributions for the sake of better visualization.}
    \label{fig:joint_posterior}
\end{figure*}
In the absence of signal detection, we move to GW strain upper limit estimation from marginalised posterior distribution. Direction of propagation, source binary orientation and phase posteriors are completely uninformative, perfectly following the prior distributions, as shown in Fig. \ref{fig:joint_posterior}. \\
Since we are studying observables proportional to the first time derivative of the GW, the amplitude of the effect is $\propto h_0 \cdot f$. In order to disentangle and properly estimate the sensitivity curve, we demote the GW frequency from being a free parameter, fix it to a specific value and run the MCMC. A robust upper limit on GW observable strain is set from the 95-th quantile of the posterior distribution. Then, the process is repeated on the whole accessible frequency range for the combined dataset. We stop the analysis at a conservative value of $f= (4T_{obs})^{-1}$ to keep the approximation $\omega t <<1 $ valid.\\
\begin{figure*}
    \centering
    \includegraphics[width=0.9\linewidth]{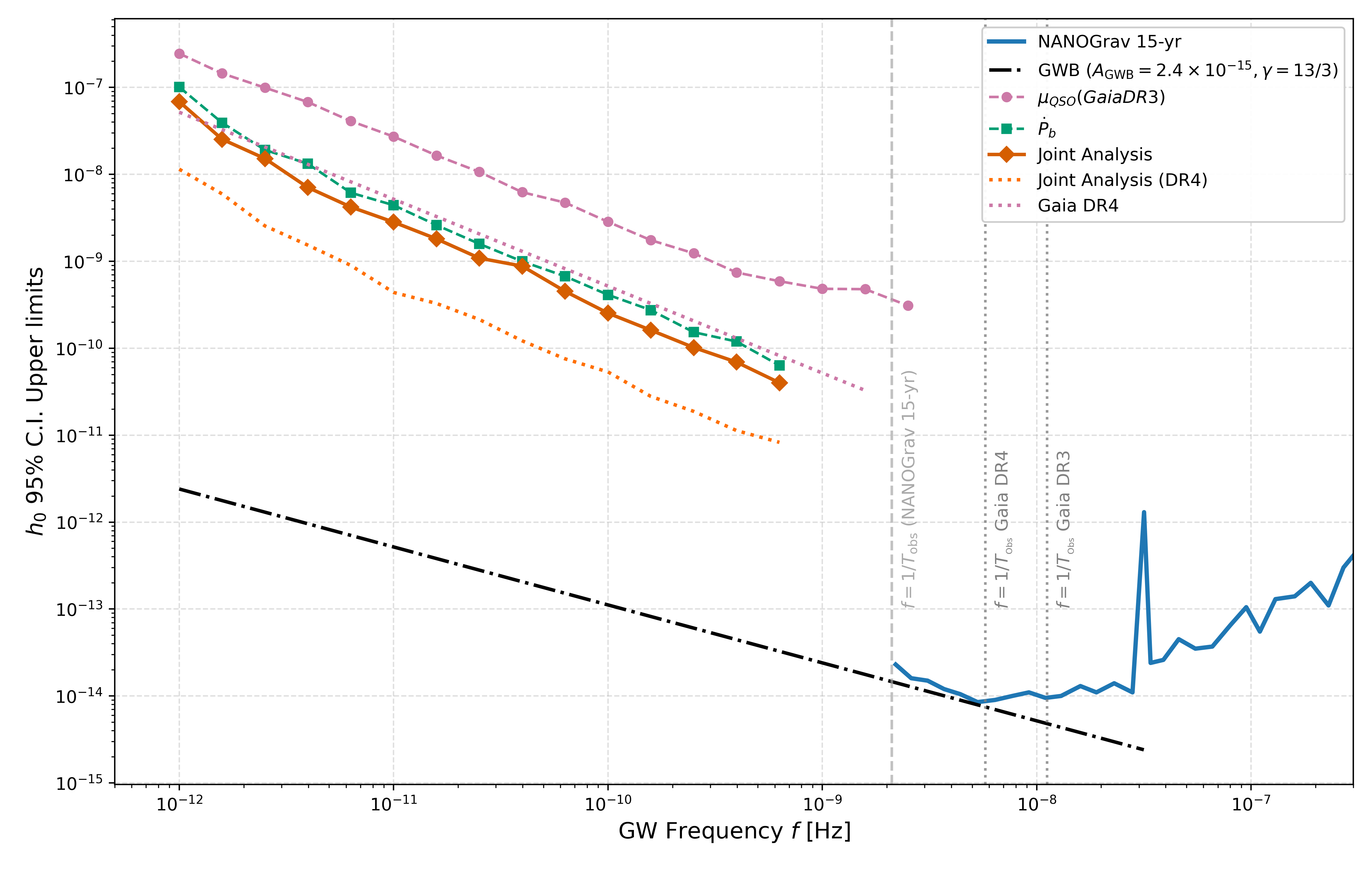}
    \caption{Upper limit estimation on CW strain for different datasets. Dashed lines are the results from the two disjoint analysis, while the orange solid lines show the improvements from the new combined method. Forecasts of results with estimated Gaia DR4 errors are also presented with dotted lines.}
    \label{fig:sens_curve}
\end{figure*}
The results are shown in Fig. \ref{fig:sens_curve}, where we find the upper limit for the join analysis to be $h_0 \leq 6.4 \times10^{-11}$ at the reference frequency of $f_{ref} = 4\times10^{-10}$ Hz. Due to current data quality, Gaia DR3 sensitivity to pHz CW is still 2-3 times less performative than the one achieved with pulsar binaries data. Hence, the combined analysis improves the upper limits by a factor of 20-30\% with respect to the analysis on $\dot P_b$ alone. Nevertheless, we achieve the tightest constraints to date on pHz CWs.  \\
\begin{figure}[h!]
    \centering
    \includegraphics[width=0.95\linewidth]{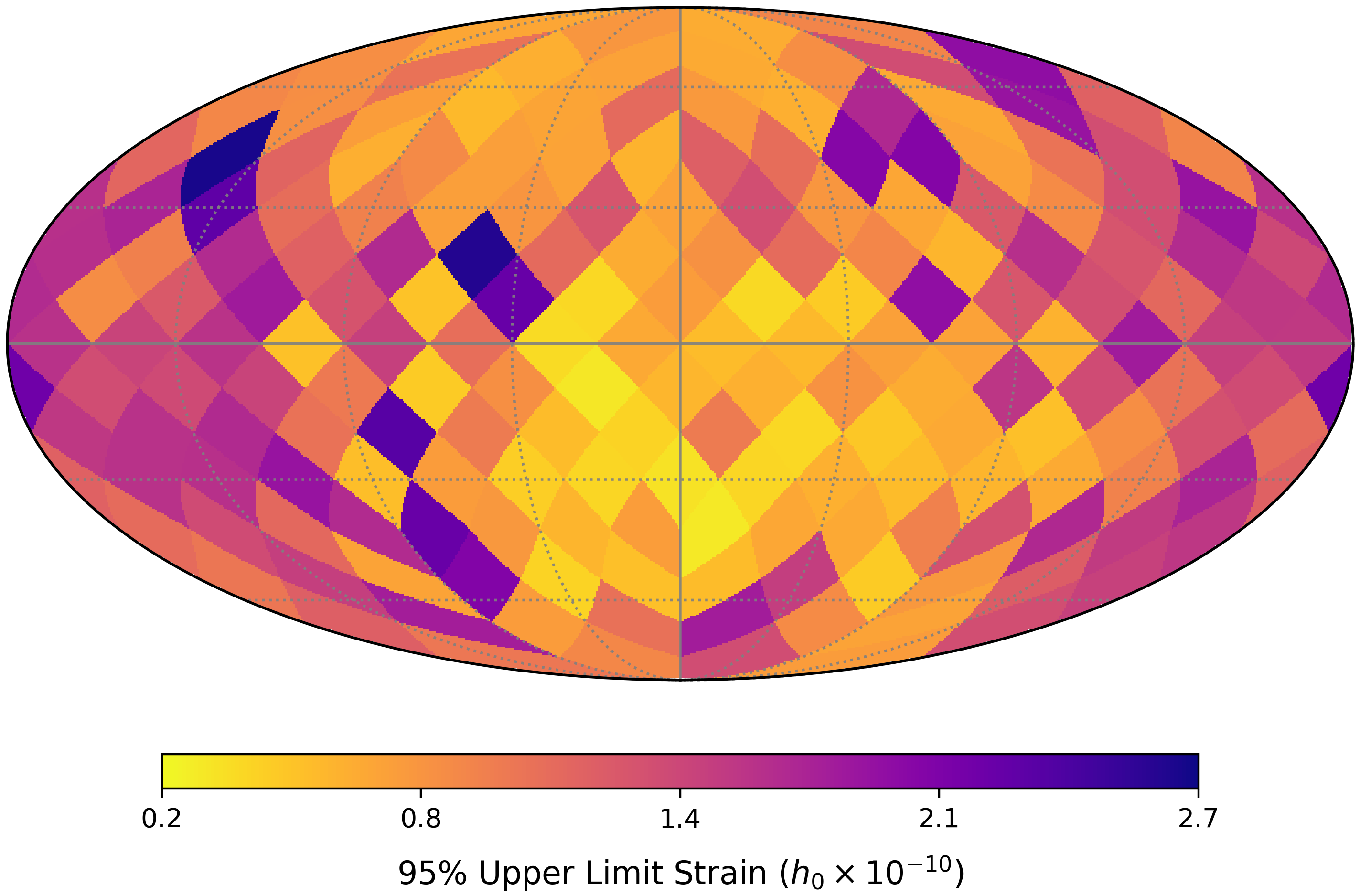}
    \caption{Directional sensitivity on CW strain for the joint datasets at a standard frequency value of $f_{GW} = 2.5 \times 10^{-10}$ Hz.}
    \label{fig:dir_sens}
\end{figure}
We also study the directional sensitivity of the combined datasets by fixing the frequency value to $f_{GW} = 2.5 \times 10^{-10}$ Hz and running the MCMC with the incoming GW direction set to a specific value. The analysis is repeated on the whole sky and the results are presented in Fig. \ref{fig:dir_sens}. The smallest achieved upper limit is $h_0 \leq 2.94 \times 10^{-11}$ while the least sensitive region yields $h_0 \leq 2.65 \times 10^{-10}$. The spatial sensitivity follows the expected antenna pattern of the chosen binaries and the spatial density of used quasars.

\begin{figure}[h]
    \centering
    \includegraphics[width=0.95\linewidth]{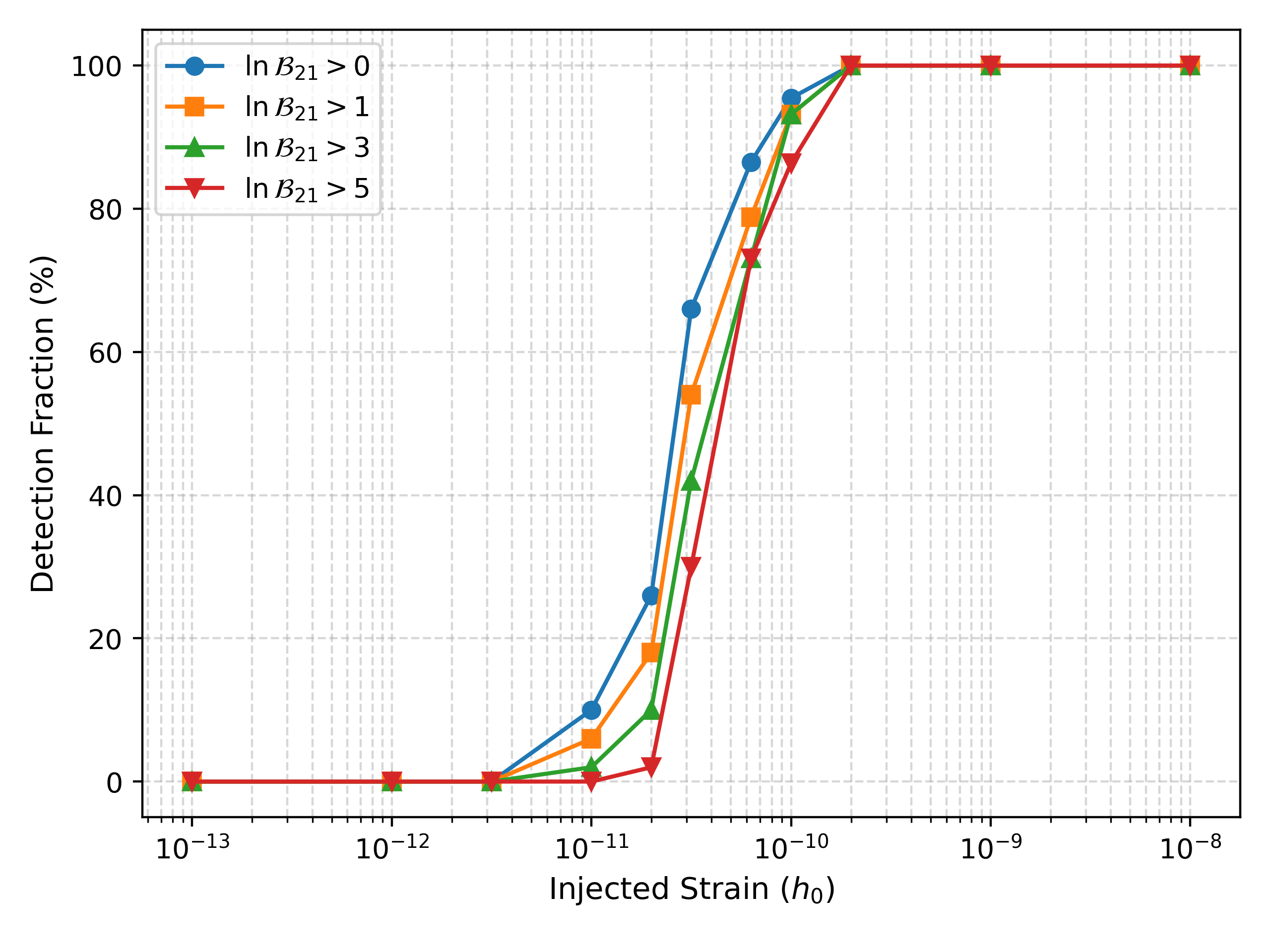}
    \caption{Retrieved Bayes factor from injected signals at reference frequency $f_{GW} = 2.5 \times 10^{-10}$ Hz. For each strain bin the percentage of BF surpassing notable threshold is shown with different colours.}
    \label{fig:inj_BF}
\end{figure}

\section{Gaia DR4 forecasts}\label{sec:future}

With the upcoming Gaia fourth data release (at the time of writing), we can forecast the possible performances of the QSO catalog and estimate the expected improvements on the CW joint analysis. \\
The main upgrade comes from the increased time baseline: while astrometric position uncertainties scale as $N_{obs}^{-1/2} \propto T^{-1/2}$, proper motion uncertainties benefit additionally from the extended temporal lever arm, scaling as
$T^{-3/2}$. Assuming a scanning rate roughly linear with time, this results in the following estimated precision:

\begin{equation}\label{eq:time_baseline}
    \sigma_{\text{DR4}} \approx \left( \dfrac{T_{\text{DR3}}}{T_\text{DR4}} \right)^{3/2}\sigma_\text{DR3} \approx 0.37 \sigma_\text{DR3} \;,
\end{equation}
as we move from a 34-month observation to 5.5 years. An additional performance improvement comes from the number of sources: in a regime of SNR effectively equal to zero, the proper motion magnitude will be proportional to the measurement uncertainty. Therefore, with better precision due to a longer time baseline, the number of sources surviving the 1 mas/yr cut will increase, ideally converging to the size of the whole catalog. Finally, in the next data release, Gaia might see new sources with full astrometric solution and the usage of epoch astrometry shall improve the purity of the dataset.

However, we only addressed the white component of the proper motion noise, while in reality a systematic floor is non negligible \citep{Lindegren2021}. Moreover, the additional sources after the proper motion cut might be characterised by worse measurement precision (high proper motion values can be correlated with fainter sources \citep{Gaia:2021gsq}), bringing down the overall sensitivity of the dataset. 

A mock catalog with real DR3 quasars position and expected new errors is constructed, and a new analysis is performed. To do so, we inject a CW signal from simulated SMBHB using \texttt{oddAGN} package. For each frequency bin, we generate 20 loud binaries with random direction in sky, in order to average the possible loss of sensitivity due to the anisotropic quasar distribution, run the analysis pipeline and take the mean upper limit. 

The same approach is repeated after creating a mock catalog of pulsar binaries with realistic astrophysical parameters and simulated values of observed $\dot P_b / P_b$; then we find the forecast joint sensitivity curve from mock DR4 data. We see that, in a fairly realistic scenario, a combined analysis might break the wall of $h_0 < 10^{-11}$ 95\% upper limit for sub-nHz CW signals. 

Finally, a injection-recovery method is used to estimate forecasts, limits of detectability, and parameter estimation performance. For each strain amplitude bin, we inject 100 mock signals with frequency set to $f_{GW} = 2.5 \times 10^{-10}$ Hz and uniformly random angular variables in order to marginalise over them. Then, the bayesian evidence is calculated for the null models and GW models to compute the log Bayes factor. As shown in Fig. \ref{fig:inj_BF}, no mock datasets showed a positive log BF below $h_0 = 10^{-11}$, while for all injections greater than $h_0=10^{-10}$ there was very strong evidence for the signal model. We conclude that with the next data release, the false positive/negative region will be constrained between these two orders of magnitude, while set $h_0=10^{-10}$ as the 95\% true positive threshold. 

\begin{figure}
    \centering
    \includegraphics[width=0.95\linewidth]{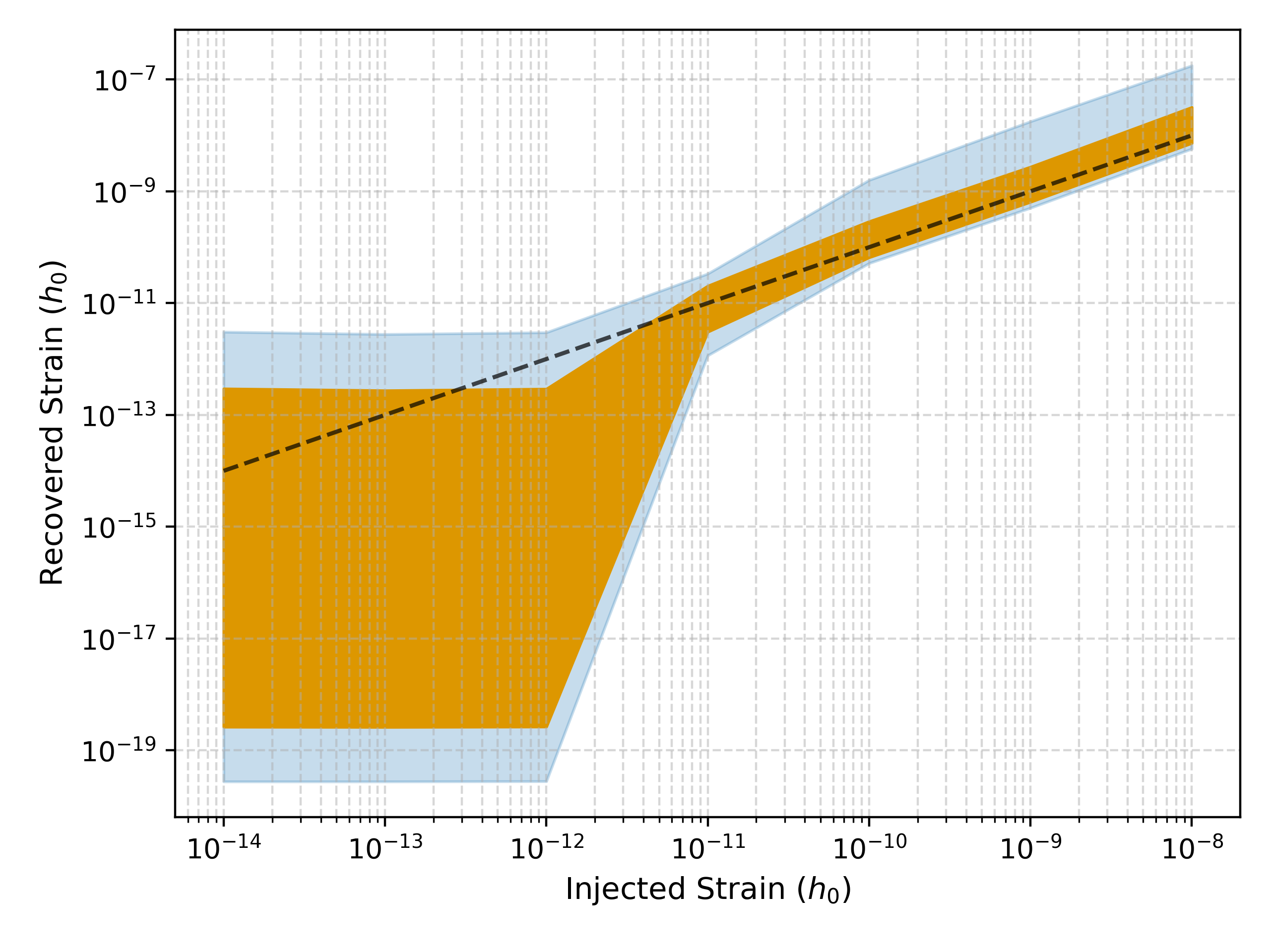}
    \caption{Strain amplitude estimation from injection of mock signals at the reference frequency of $f_{GW} = 2.5 \times 10^{-10}$ Hz. The orange shaded area represents the 68\% confidence interval (i.e. $1\sigma$ region from the best-fit parameters), while the blue shaded area is the 95\% confidence interval.}
    \label{fig:inj_rec}
\end{figure}

In Fig. \ref{fig:inj_rec} we present the retrieved strain from the same injection recovery approach. It clearly shows that in the strain region of detectability the algorithm is perfectly estimating the parameters, while a noticeable loss of precision starts to kick in below $h_0=10^{-11}$, in perfect accordance with the previous plot.

Moreover, Gaia DR4 will also publish single epoch astrometric and photometric time series. Such observables will open up new frontiers for the first searches of GW periodic signals with astrometry, with possibilities of a full cross-correlation from astrometry and pulsar timing analysis, spanning from pHz to mHz and covering more than 5 orders of magnitudes.

\section{Conclusions}
In this work, we have presented the first joint data analysis pipeline combining real astrometric measurements from Gaia CRF3 proper motions with orbital period derivatives ($\dot{P}_b/P_b$) of binary pulsars from the ATNF catalog, for searching and constraining CWs in the ultra-low frequency ($\text{pHz}$) regime ($10^{-12}\,\text{Hz} \lesssim f_{\text{GW}} \lesssim 10^{-9}\,\text{Hz}$). This hybrid approach takes advantage of the deep geometrical complementarity of the two observational techniques: pulsar binary timing is sensitive to the line-of-sight acceleration induced by passing GWs, whereas astrometric measurements of distant quasars detect the transverse sky-plane apparent proper motion displacement. Thanks to the combination of these complementary effects, the joint framework helps break the spatial degeneracies inherent to single-method searches and provides complementary sensitivity to the full three-dimensional GW response.

Employing Bayesian model selection via Dynamic Nested Sampling and the Savage-Dickey Density Ratio (SDDR), we found no statistically significant evidence for a GWs signal in either the isolated datasets or the joint analysis ($\ln \mathcal{B}_{\text{joint}} = -0.42 \pm 0.03$). 

In addition to that, accounting for systematic effects proved critical to avoid spurious strain signals. For pulsar binaries, incorporating Monte Carlo propagation of Galactic potential parameters provided the necessary variance to absorb unmodeled residual power. For quasars, modelling and implementing systematic correlations through Vector Spherical Harmonics up to the octupole order ($\ell = 3$) prevented artificially inflated GW strain from parameter estimation. \\
In the absence of a detection, we derived the tightest upper limits to date on CW strain in the $\text{pHz}$ band. For a reference frequency of $f_{\text{ref}} = 4 \times 10^{-10}\,\text{Hz}$, the joint analysis resulted in a $95\%$ upper limit of $h_0 \le 6.4 \times 10^{-11}$, representing an improvement of $20\%-30\%$ over the limits obtained from binary pulsar timing alone. Directional sensitivity maps yield upper limits ranging from $h_0 \le 2.94 \times 10^{-11}$ in the most sensitive sky directions to $h_0 \le 2.65 \times 10^{-10}$ in less constrained regions, also improving the performance and scattering in directional sensitivity compared to single-datasets analysis.

Looking ahead, several developments will substantially extend the sensitivity and scope of this framework, with Gaia DR4 as the closest and primary upgrade. With an extended observational baseline of 5.5 years, individual proper motion uncertainties are expected to decrease significantly. Our injection-recovery forecasts demonstrate that a joint analysis with Gaia DR4 can break the $h_0 < 10^{-11}$ upper limit barrier for sub-$\text{nHz}$ CWs. Furthermore, the publication of single-epoch astrometric time series in DR4 will enable direct searches for periodic GW signals spanning from $\text{pHz}$ to $\text{mHz}$. In addition to that, the arrival of the Square Kilometre Array will dramatically enhance the number of observed millisecond binary pulsars, particularly toward the Galactic bulge \citep{SKAOPulsarScienceWorkingGroup:2025syv}. A larger and more isotropically distributed sample of high-precision pulsar binaries will substantially improve  the directional coverage and three-dimensional sensitivity of the joint analysis.

Future implementations of this pipeline will incorporate other secular observables, such as pulsar spinning period jerks (i.e. $\ddot{P}$), since cross-correlation of proper motions with a second order time derivative could break strain-frequency degeneracies. Finally, extending this joint framework to searches for a stochastic gravitational-wave background (SGWB) could provide new opportunities to probe cosmological sources in the $\text{pHz}-\text{nHz}$  frequency range, including possible signatures of early-Universe phase transitions.

\begin{acknowledgements}
Data from the European Space Agency (ESA) mission Gaia (\url{https://www.cosmos.esa.int/gaia}), processed by the Gaia Data Processing and Analysis Consortium (DPAC, \url{https://www.cosmos.esa.int/web/gaia/dpac/consortium}) was used for this work, along with ATNF (\url{https://www.atnf.csiro.au/research/pulsar} for the pulsar analysis. L.F., M.C., acknowledge support from the PRIN 2022 grant 20227MYL2X “General Relativistic Astrometry and Pulsar Experiment (GRAPE) financed by the European Union - Next Generation EU, Mission 4 Component 1 CUP C53D23000890006. \\
\end{acknowledgements}

\bibliographystyle{aa_1}
\bibliography{bibliography}

\begin{appendix}

\onecolumn
\section{Uncertainty on the Galactic acceleration}\label{sec:gal_model}
Since distances are free parameters in the MCMC, the Galactic contribution is marginalised only over the distances, while RA and DEC are fixed, as their relative uncertainties are orders of magnitude below the parallax average error.
However, due to our strict cut on the parallax and parallax error, the change in the $a_{gal}$ term is much lower than the instrumental error for most of the binaries, as shown in Table \ref{tab:gal_accelerations}. This results in an underestimation of the total uncertainty for the expected Galactic contribution, as discussed in \citep{Moran:2023myv}. However, for our work we do not implement a jitter noise parameter to deal with this excessive power, but we implement the model uncertainty in the analysis with a Monte Carlo approach. Since \citet{McMillan2017} is the sum of different contributions (bulge, gas disk, thin and thick stellar discs, HI and molecular gas discs, and a DM halo, for a total of 7 free parameters. In order to evaluate the uncertainty on the acceleration, for each parameter, we consider the mean and standard deviation obtained from the posterior distribution \citep[see Table 2 of][]{McMillan2017}, with the exception of the bulge central density that we keep fixed to the best-fitting value.
From these values, we define Gaussian distributions (truncated to avoid negative values) from which we draw 10k random values for each of the 6 free parameters. \\
Then, from each simulated vector of free parameters, we evaluate the Galactic accelerations for all pulsar binaries and calculate the corresponding standard deviations. For each pulsar binary, we adopt this standard deviation as an estimator for the model-induced uncertainty on the Galactic acceleration. \\
After implementing this additional error in the likelihood, the issue of spurious detection is no longer present, indicating a possible underestimation of the $\dot P_b$ errors in the catalog or an inadequate understanding of the local Galactic potential and its modelling. Improving Galactic models is outside the scope of this work, but it will be an interesting challenge for the near future, in light of the next Gaia data release. \\
A combined analysis to constrain the (local) Galactic potential could bring important advantages: since we can only observe pulsars' line of sight acceleration, a strong constraint on the Solar System acceleration can, in principle, disentangle this problem and let us study the proper Galactic acceleration acting on nearby pulsar binaries.

\begin{table*}[hb!]
    \centering
    \begin{tabular}{c|c|c|c}
    \hline \hline
 \toprule
 \textbf{ID} & $- 5\sigma$ & Best-fit & $5 \sigma$ \\
 \midrule
 \hline
     J0437-4715 & -0.048 & -0.049 & -0.051 \\
J0613-0200& 0.023 &0.032 &0.043 \\
J1012+5307 &-0.080 &-0.080& -0.079\\
J1022+1001 &-0.103& -0.118& -0.125\\
J1455-3330 &-0.022 &-0.031 &-0.041\\
J1518+4904 &-0.152 &-0.167& -0.180\\
B1534+12 &-0.115 &-0.146 &-0.175\\
J1614-2230& 0.002 &0.006 &0.010\\
J2145-0750 &-0.112 &-0.138& -0.162\\
J2222-0137 &-0.083 &-0.084 &-0.085\\
J2234+0611 &-0.162 &-0.184 &-0.205\\
\bottomrule
    \end{tabular}
    \caption{Expected Galactic accelerations (in $10^{-18} s^{-1}$ units) from McMillan17 for the selected binaries at different distance values. We show $a_{gal}$ at $d_p \pm 5\sigma$ and the best-fit value.}
    \label{tab:gal_accelerations}
\end{table*}

\FloatBarrier 
\twocolumn

\begin{table*}[ht!]
\section{Pulsar Binaries data and formulas}
Taking the time derivative of eq. \eqref{eq:gw_velocity} one gets the following analytical formula for the GW-induced acceleration:
\begin{equation}\label{eq:gw_acc}
\begin{aligned}
a_{GW} &= \frac{\hat{n}_p^i \hat{n}_p^j}{1 + \hat{k} \cdot \hat{n}_p} \omega_0 h_0 [ e_{ij}^+(\hat{k}) [\cos(2\Psi)(1 + \cos^2(i))(\cos(2\phi_0) - \cos(2\phi_0 - 2\omega_0(1 + \hat{k} \cdot \hat{n}_p)d_p)) + \\
&+ 2 \sin(2\Psi) \cos(i)(-\sin(2\phi_0) + \sin(2\phi_0 - 2\omega_0(1 + \hat{k} \cdot \hat{n}_p)d_p))] +\\
&+ e_{ij}^\times(\hat{k}) [\sin(2\Psi)(1 + \cos^2(i))(\cos(2\phi_0) - \cos(2\phi_0 - 2\omega_0(1 + \hat{k} \cdot \hat{n}_p)d_p)) +\\
&- 2 \cos(2\Psi) \cos(i)(-\sin(2\phi_0) + \sin(2\phi_0 - 2\omega_0(1 + \hat{k} \cdot \hat{n}_p)d_p))]] \;,
\end{aligned}
\end{equation}
under the approximation $\omega t \ll 1$, i.e. the epoch of observation is negligible compared to the period of the wave. \\
We now provide the pulsar binaries dataset used for the analysis: \\

\label{table:pta_astro} 
\centering
            
\begin{tabular}{c|c|c|c|c|c|c}
  \hline\hline 

\toprule
\textbf{ID} &  $\varpi$ [mas] & $\alpha$ [deg] & $\delta$ [deg] & $P_b$ [d]  & $\dot P_b$  & Ref.\\
\midrule
\hline
J0437-4715 	&	 6.43(4)   	&	69.317	&	-47.253	& 5.7410459(4)   &   	 $3.7329(15) \times 10^{-12}$ & \citep{EPTA:2023sfo} \\
J0613-0200 &	 	1.0(5)  &  		93.433	&	-2.013	& 1.198512575192(8) &	 $3.5(2) \times 10^{-14}$ & \citep{EPTA:2023sfo}    \\
J1012+5307 	 &	1.17(2)   	&	153.139	&	53.117&	 0.604672722921(3) &	 $5.46(6) \times 10^{-14}$ & \citep{EPTA:2023sfo}   \\
J1022+1001 	&	 1.16(8) 	 & 	155.742	&	10.031&	 7.805134(2)       &	 $2.2(1) \times 10^{-13}$  & \citep{EPTA:2023sfo}   \\
J1455-3330 	&	 1.3(1)  	  &	223.956	&	-33.513&	 76.174567479(8)&   	 $5(2) \times 10^{-12}$    &\citep{EPTA:2023sfo}   \\
J1518+4904	 &	 1.24(4)	  & 	229.57	&	49.076&	 8.6340050964(11)&  	 $2.4(22) \times 10^{-13}$  & \citep{Janssen:2008mh}  \\
B1534+12   	&	 1.07(7) 	 & 	234.292&		11.932&	 0.420737298879(2)& 	 $-1.366(3) \times 10^{-13}$ & \citep{Fonseca:2014qla} \\
J1614-2230 	&	 1.54(1)  	 &	243.652	&	-22.509	 &8.68661942256(6)  	& $1.57(13) \times 10^{-12}$  & \citep{NANOGrav:2020gpb} \\
J2145-0750 	&	 1.4(1)   	 &	326.46	&	-7.838	 &6.83890261495(9)  &	 $1.3(2) \times 10^{-13}$  & \citep{Reardon:2021gko}   \\
J2222-0137 	&	 3.723(13) 	&	335.525	&	-1.621	& 2.445759995471(6) &	 $2.55(8) \times 10^{-13}$ & \citep{Guo:2021bqa}   \\
J2234+0611 	&	 1.03(4)   	&	338.596	&	6.191	& 32.00140163(8)    &	 $3.1(25) \times 10^{-12}$ & \citep{Stovall:2018rvy} \\

\bottomrule
\hline
\end{tabular}
\caption {Astrometric and timing parameters of pulsar binaries used in the main analysis, including sky position in ICRS, parallax, binary period and its first time derivative. The data were obtained from the publicly available ATNF catalog.\citep{Hobbs:2003gk}.} 
\end{table*}

\begin{figure*}
    \centering
    \includegraphics[width=0.95\linewidth]{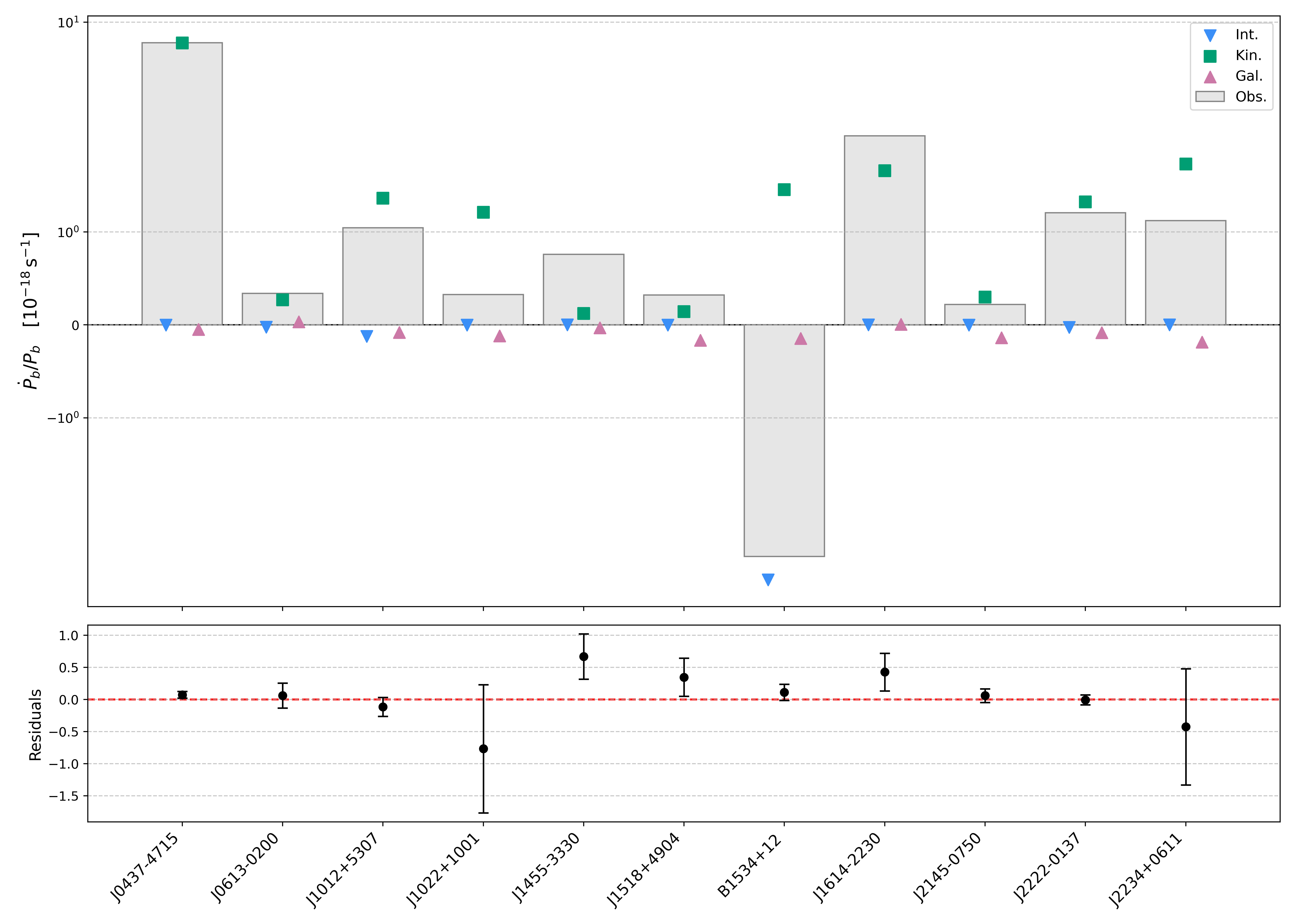}
    \caption{Observed values of fractional change of binary periods and their expected contributions. Intrinsic GW emission is shown in blue, while Galactic expected acceleration in pink and Shklovskii contribution in green. At the bottom, the final residuals used for the analysis with corresponding uncertainties. }
    \label{fig:pta_contributions}
\end{figure*}

\FloatBarrier 
\twocolumn

\onecolumn
\section{Gaia systematics and Harmonic Decomposition}\label{sec:gaia_systematics}
As explained in Sec. \ref{sec:theo_frame}, the Gaia CRF3 catalog is heavily affected by systematic effects, some of which have a clear astrophysical origin, while others are just instrumental. A very useful and elegant approach for addressing this issue is the implementation of a harmonic decomposition using VSH functions. With this method, we can decompose a two-dimensional vector field on the sphere as a sum of E-mode and B-mode bases (i.e. pure divergence and pure curl vector bases) \citep{Mignard:2012xm}:
\begin{align}
    \vec{Y}_{\ell m}^E &= \frac{1}{\sqrt{\ell(\ell + 1) }} \vec{\nabla} Y_{\ell m} (\alpha, \delta) \;, \\
    \vec{Y}_{\ell m}^B &= -\frac{1}{\sqrt{\ell(\ell + 1) }} \mathbf{u} \times \vec{\nabla} Y_{\ell m} (\alpha, \delta) \;,
\end{align}
where the superscripts $E$ and $B$ correspond to the spheroidal (electric-like) and toroidal (magnetic-like) modes, respectively, while $Y_{\ell m}$ is the standard spherical harmonic function. A generic two-dimensional vector on the sphere will be 
\begin{equation}
    \vec{V} (\alpha, \delta) = \sum_{\ell = 1}^{\infty} \sum_{m = - \ell} ^\ell (s_{\ell m} \vec{Y}_{\ell m}^E (\alpha, \delta) + t_{\ell m} \vec{Y}_{\ell m}^B (\alpha, \delta)) \;,
\end{equation}
where $s_{\ell m}$ and $t_{\ell m}$ are complex expansion coefficients describing the amplitudes of the corresponding E- and B-mode multipoles.

\begin{figure*}
    \includegraphics[width=0.5\linewidth]{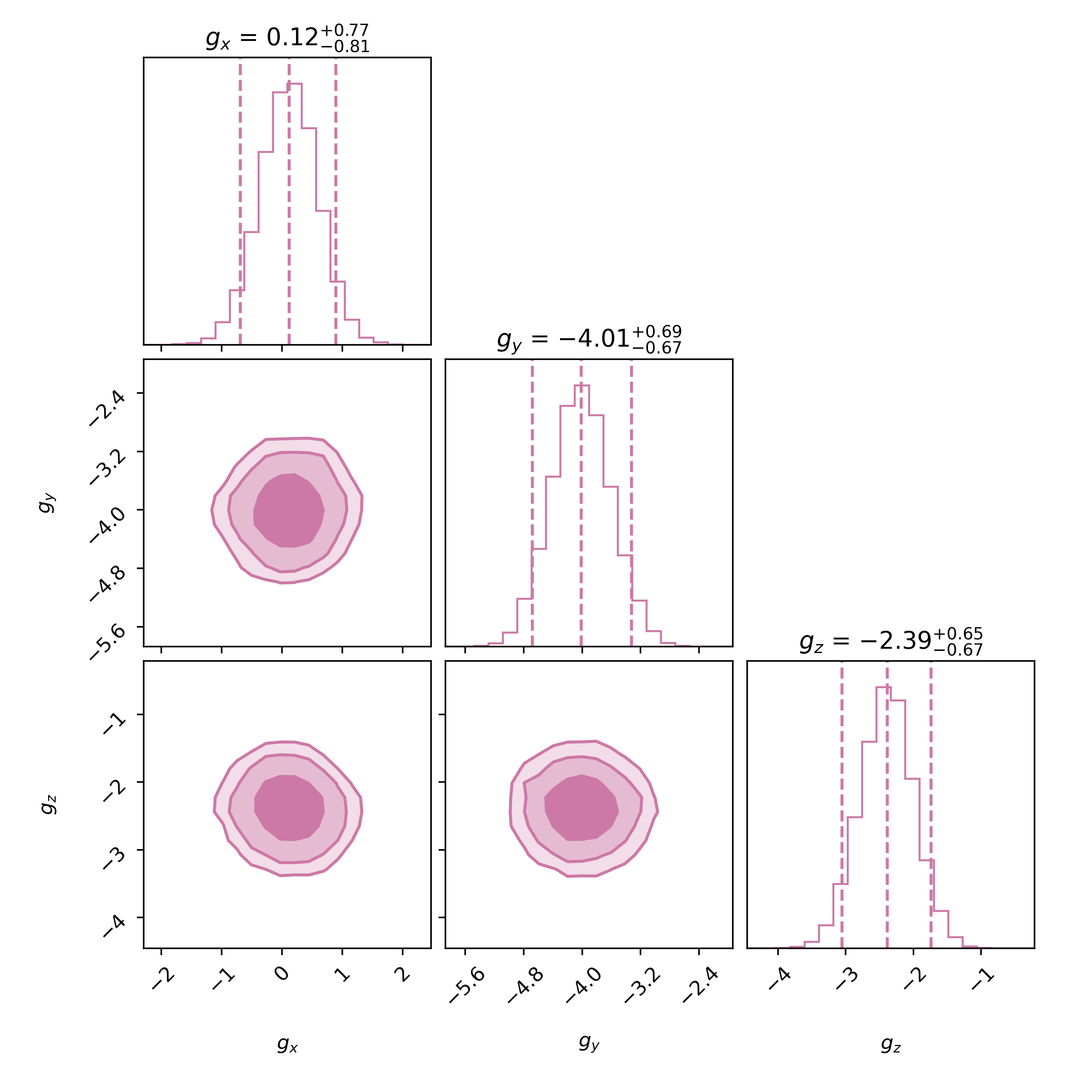}
    \includegraphics[width=0.5\linewidth]{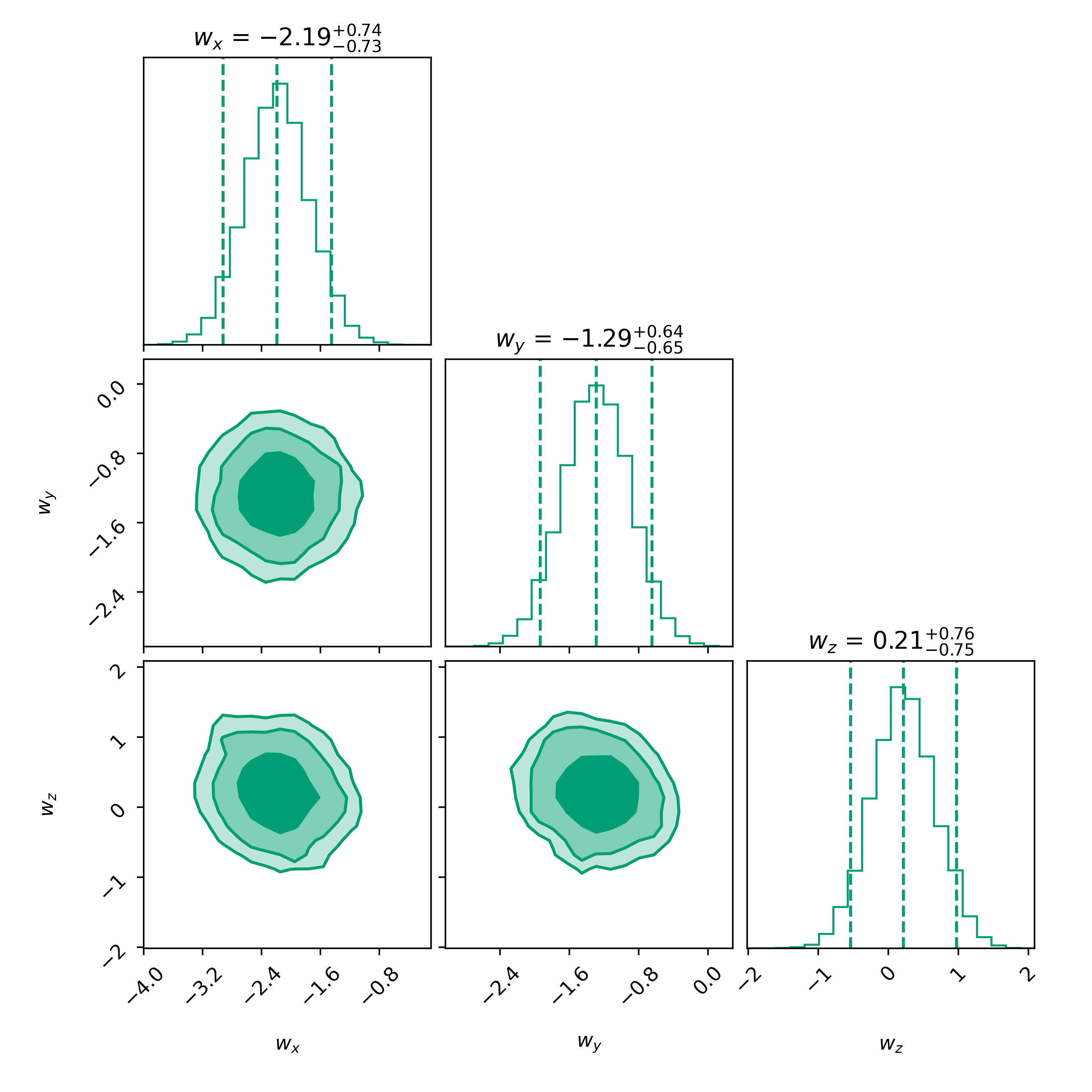}
    
    \caption{Dipole contribution extracted from the MCMC analysis of the Gaia CRF3 dataset. Left: posterior distributions of secular acceleration vector components. Right: posterior distributions of rotation vector components.}
    \label{fig:dipole}
\end{figure*}

\begin{figure*}
    \includegraphics[width=0.5\linewidth]{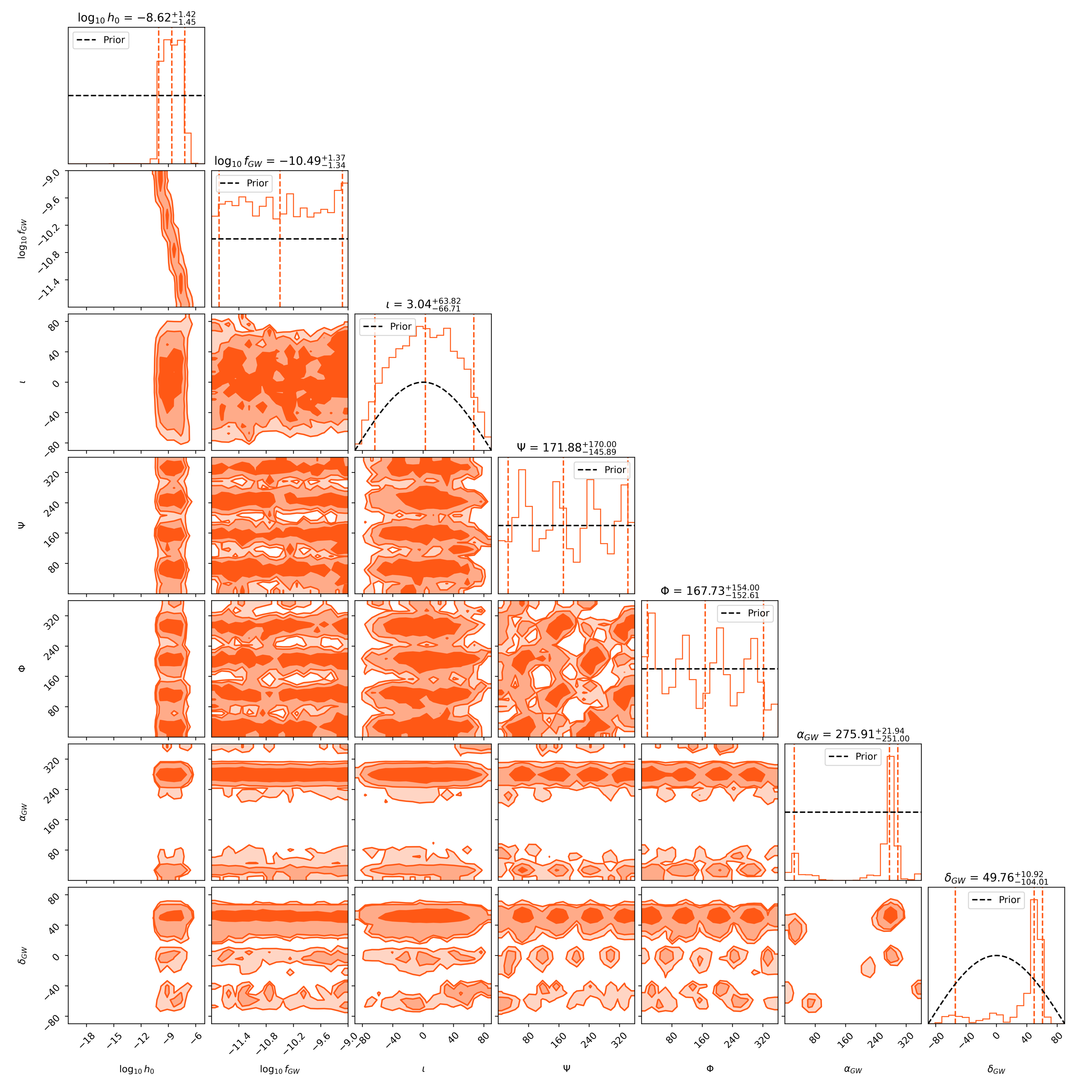}
    \includegraphics[width=0.5\linewidth]{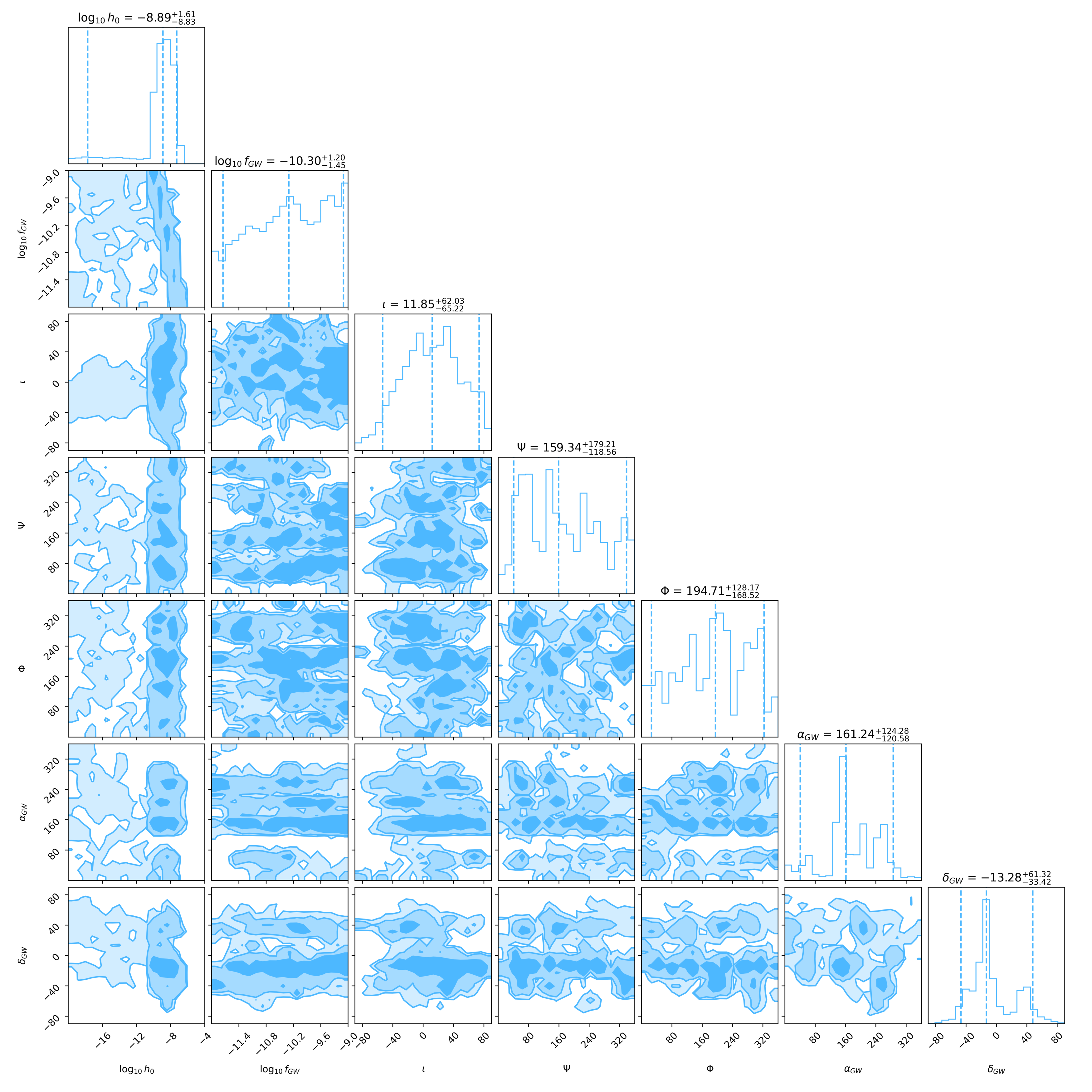}
    
    \caption{Left: posterior distributions of GW parameters from the Gaia CRF3 assuming only dipole contributions. Right: posterior distributions from the same dataset after including generic quadrupole contributions.}
    \label{fig:qso_issues}
\end{figure*}

In this formalism, two important contributions to quasars proper motion (i.e. secular aberration, or glide, and rotation of the reference system and/or the universe) can be interpreted as pure dipole $\ell = 1$ terms, while the main contribution coming from a passing GW (CW or Stochastic Background) is confined to the quadrupole $\ell =2$ terms, and the remaining power leaks to smaller scales with amplitude $\propto (1 + \hat k \cdot \hat n)^{-\ell}$ for each multipole.
Rotation and glide will produce an apparent proper motion with the following characteristics:
\begin{equation}
\begin{aligned}
\mu_\alpha^* &= \left( -g_x \sin\alpha + g_y \cos\alpha \right) + \left( -w_x \cos\alpha \sin\delta - w_y \sin\alpha \sin\delta + w_z \cos\delta \right) \;, \\
\mu_\delta &= \left( -g_x \cos\alpha \sin\delta - g_y \sin\alpha \sin\delta + g_z \cos\delta \right) + \left( w_x \sin\alpha - w_y \cos\alpha \right) \;, \\
\end{aligned}
\end{equation}
where $\vec g = (g_x, g_y, g_z)$ is the Solar System acceleration vector (divided by $c$) and $\vec \omega = (\omega_x, \omega_y, \omega_z)$ measures the non-inertiality of the reference system or the rotation of the universe at cosmological scales \citep{Akhmetov:2021}. In Fig. \ref{fig:dipole}, we show the posterior distributions for the glide and rotation contributions from our analysis, while the prior distributions are described in Sec. \ref{sec:data_analysis}. Our retrieved secular aberration has magnitude compatible with the Gaia CRF3 reference value within $1 \sigma$, with direction pointing towards the Galactic centre within the error range (i.e. $\alpha_g = 271.71 \degree$ and $\delta_g = -30.78 \degree$). \\
\begin{figure*}
    \centering
    \includegraphics[width=0.8\linewidth]{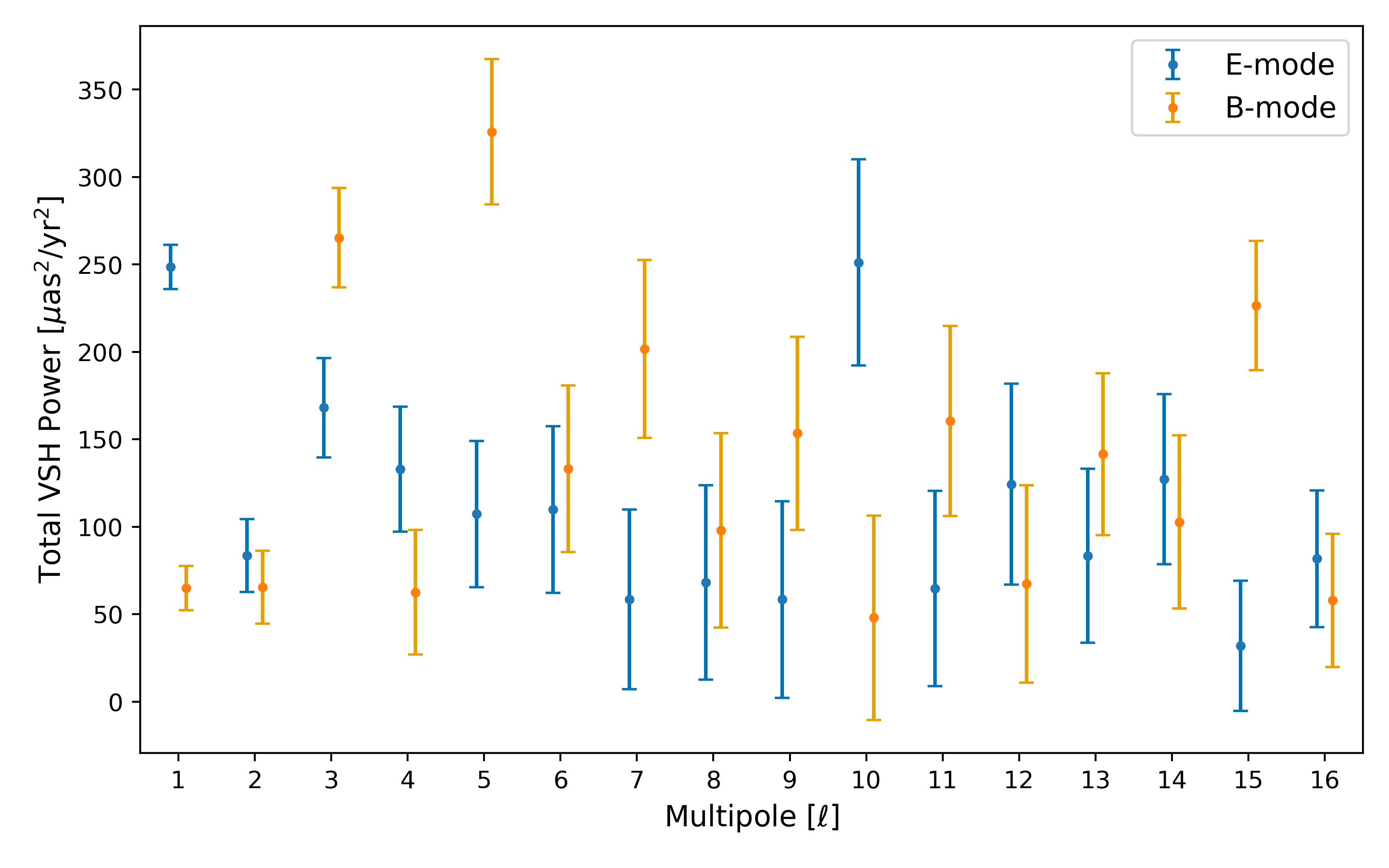}
    \caption{VSH total power of the Gaia CRF3 catalog at different multipoles. The blue points represent the total power of the pure divergence vector base, while the pure curl vector base is displayed in orange. The offset on the x-axis is implemented for better visualization.}
    \label{fig:VSH_power} 
\end{figure*} 
As seen in Fig. \ref{fig:VSH_power}, higher multipoles have non-negligible power, indicating the presence of systematics at medium/small scales. A CW analysis is impossible to perform on the CRF3 catalog without accounting for these contributions in the residuals calculation (or without performing a strict cut on the proper motion amplitude in order to discard all sources with significant systematic contributions, but that would bring to underestimation bias as shown in \citep{Akhmetov:2026bfn}). We show in Fig. \ref{fig:qso_issues} that with a model comprehending only glide, rotation and CW, the fitting algorithms forces the wave amplitude to non-physical values to explain all the remaining large-medium correlations in proper motion. Therefore, we decide to include in our model also generic systematic effects using VSH decomposition up to $\ell=3$, as this was the best compromise between physical results and computational cost. \\
In principle, the VSH coefficients could absorb the GW signal, as the total quadrupole VSH power is the standard proxy in astrometry for the detection of a GWB \citep{Mignard:2012xm}. However, in the case of a CW, only seven parameters are necessary for describing the phenomenon, meaning that the VSH coefficients must take an extremely specific configuration to eliminate the extra degrees of freedom (ten coefficients from quadrupole contribution plus GW power leakage at higher multipole orders). As a result, due to the lower dimensionality of the model, in the presence of a passing CW, the fitting algorithm will prefer the GW model and pass the remaining unmodeled quadrupole signal to the VSH coefficients.

\begin{figure*}
    \includegraphics[width=0.95\linewidth]{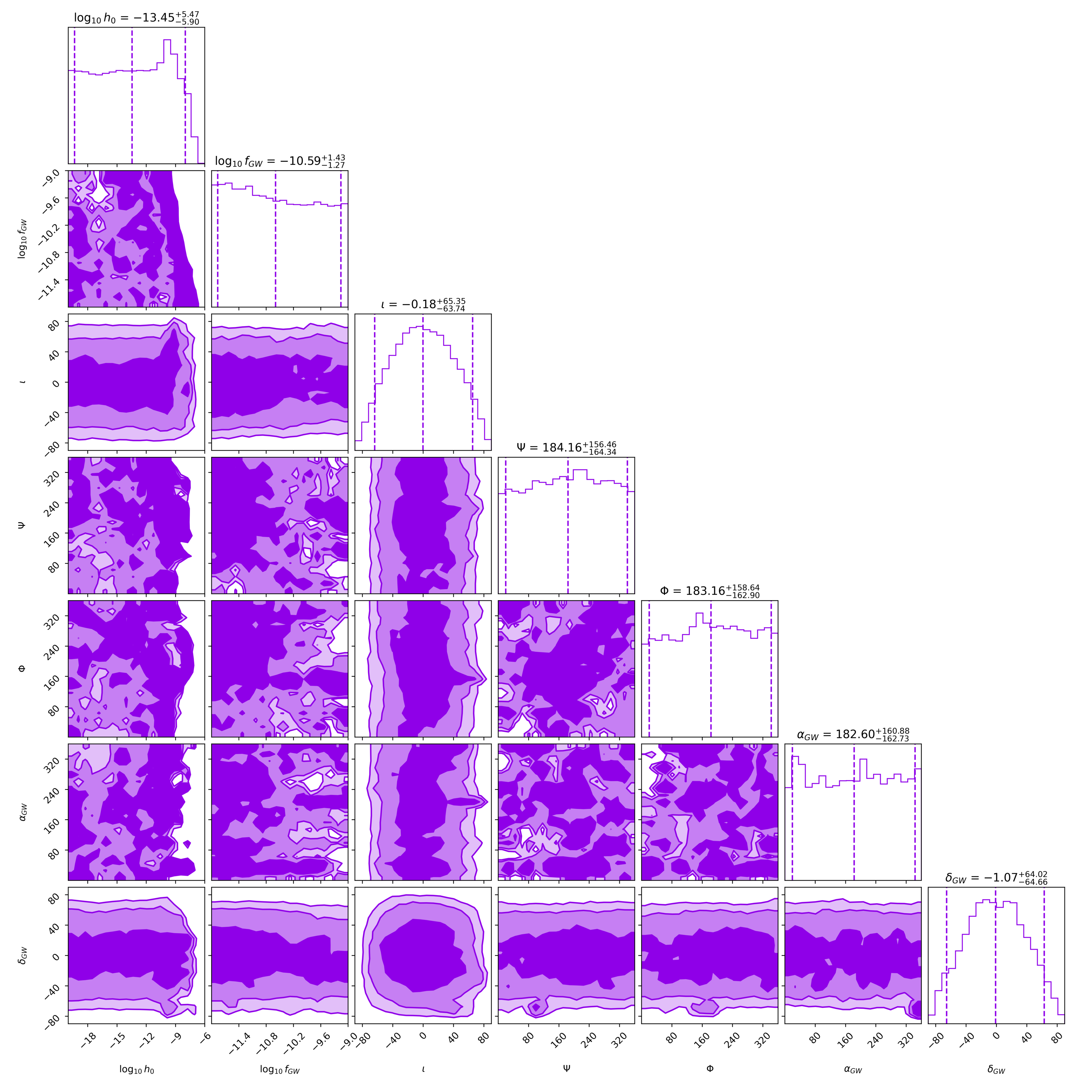}
    
    \caption{Posterior distributions of GW parameters after implementing VSH coefficients as free parameters up to the order $\ell =3$.}
    \label{fig:qso_mcmc}
\end{figure*}

\end{appendix}
\end{document}